\documentclass{aa2}  

\usepackage{graphicx}
\usepackage{color}
\usepackage{xcolor}
\usepackage{txfonts}
\begin{document} 

\def\fref{\stackrel{\hspace{0.5ex}{\scriptscriptstyle\circ}}{f}
            \hspace{-0.8ex}}
\def\tauref{\tau^{\hspace{-0.9ex}^{\circ}}}
\def\nfref{n_{\rm f}^{\hspace{-0.9ex}^{\circ}}}
\def\nrref{n_{\rm r}^{\hspace{-0.9ex}^{\circ}}}
\def\nref{n^{\hspace{-0.9ex}^{\circ}}}
\def\vref{v^{\hspace{-0.9ex}^{\circ}}}
\def\alpharef{\alpha^{\hspace{-0.9ex}^{\circ}}}
\def\pref{p^{\hspace{-0.9ex}^{\circ}}}
\def\st{{\rm \hspace{0.1ex}\circ\hspace{-1.2ex}-}}
\def\dG{\Delta_{\rm f}^{\st}\hspace{-0.2ex}G}
\def\pst{p^{\st}}
\def\Vl{{V_{\ell}}}
\def\pabl#1#2{\frac{{\rm\partial} #1}{{\rm\partial} #2}}
\def\vgas{\vec{v}_{\rm gas}}
\def\vdreq{v^{\hspace{-0.8ex}^{\circ}}_{\rm dr}}
\def\xx{\vec{x}}
\def\ie{i.\,e.\ }
\def\eg{e.\,g.\ }
\def\DV{\Delta V_r}
\def\Vs{{V_{\rm s}}}
\def\Sr{S_{\!r}}
\def\vdrn{\vec{v}_{\rm dr}^{\hspace{-0.9ex}^{\circ}}}
\def\nr0{n_{r}^{\hspace{-0.9ex}^{\circ}}}
\def\jdiff{\vec{j}^{\rm diff}_i}
\def\er{\vec{e}_r}
\def\rhod{\rho_{\rm d}}
\def\nH{n_{\langle{\rm H}\rangle}}

   \title{Dust in brown dwarfs and extra-solar planets}

   \subtitle{VIII. TiO$_2$ seed formation: 3D Monte Carlo versus kinetic approach}
   
   \titlerunning{Dust in brown dwarfs and extra-solar planets VIII. TiO$_2$ seed formation}
   \authorrunning{K\"ohn, Helling, B{\o}dker Enghoff et al. }

   \author{C. K\"ohn,
          \inst{1}
          Ch. Helling,
          \inst{2,3,4}
          M. B{\o}dker Enghoff,
          \inst{1},
          K. Haynes,
          \inst{2,3},
          D. Krog \inst{1},
          J.P. Sindel
          \inst{2,3,5},
          D. Gobrecht \inst{5},
          }
 \institute{National Space Institute (DTU Space), Technical University of Denmark, Kgs Lyngby, Denmark\\
            \email{koehn@space.dtu.dk}
           \and
Centre for Exoplanet Science, University of St Andrews, North Haugh, St Andrews, KY169SS, UK
            \and
SUPA, School of Physics \& Astronomy, University of St Andrews, North Haugh, St Andrews, KY169SS, UK 
             \and
SRON Netherlands Institute for Space Research, Sorbonnelaan 2, 3584 CA Utrecht, NL
            \and
Institute of Astronomy, KU Leuven, B-3001, Leuven, BE
}

   \date{Received ...; accepted ...}

 
  \abstract
   {{Modelling the formation of cloud condensation nuclei (CCNs) is key for predicting cloud properties in planet and brown dwarf atmospheres.}
   The large diversity of exoplanets (rocky planets, mini-Neptunes, giant gas planets) 
   requires a fundamental approach to cloud formation modelling in order to allow a full analysis of observational data contributing to exoplanet characterisation.}
   {
    We aim to understand the onset of cloud formation {and study the formation of} TiO$_2$-CCNs.  The formation of (TiO$_2$)$_{\rm N}$ clusters  as precursors to extrasolar cloud formation is modelled by two different methods in order to understand their potential, identify underlying shortcomings, and to validate our methods. We propose potential spectral tracers for TiO$_2$-CCN formation.}
   {
   We applied three-dimensional Monte Carlo (3D MC) simulations  to model the collision-induced growth of TiO$_2$-molecules to (TiO$_2$)$_{\rm N}$-clusters in the free molecular flow regime of an atmospheric gas. We derived individual, time-dependent (TiO$_2$)$_{\rm N}$ cluster number densities.
   For $T=1000$K, the results are compared to a kinetic approach that utilises thermodynamic data for individual (TiO$_2$)$_{\rm N}$ clusters.}
   { The {(TiO$_2$)$_{\rm N}$}  cluster size distribution is temperature dependent and evolves in  time until a steady state is reached.
   For $T=1000$K, the 3D MC and the kinetic approach agree well regarding the cluster number densities for $N=1\,\ldots\,10$, the vivid onset of cluster formation, and the long transition into a steady state. Collision-induced growth and evaporation simulated using a 3D MC approach enables a faster onset of cluster growth through nucleation bursts. {Different size distributions develop for monomer-cluster and for cluster-cluster growth, with the largest clusters appearing for cluster-cluster growth.}}
  {The (TiO$_2$)$_N$  cluster growth efficiency has a sweet-spot temperature at $\approx 1000$K at which CCN formation is triggered. 
  The combination of local thermodynamic conditions and chemical processes  therefore determines  CCN formation efficiency. The onset of cloud formation may be observable through the (TiO$_2$)$_4$, (TiO$_2$)$_5$, and (TiO$_2$)$_6$ vibrational lines, which may be detectable with the Mid-Infrared Instrument on the James Webb Space Telescope or the Extremely Large Telescope's mid-IR imager, but more complete line-list data are desirable.}


   \maketitle

\section{Introduction}
Cloud formation has emerged as one of the major obstacles to fully benefitting from observations of exoplanetary and brown dwarf atmospheres and efforts are therefore  gearing up to understand fundamental processes in enough detail to allow the development of efficient model descriptions (\citealt{2020arXiv201103302H}). Exoplanet cloud modelling approaches include a variety of parametrizations from imposing cloud particle properties as an opacity source (e.g. \citealt{2020arXiv201006936R,2020MNRAS.tmp.3310P}) and adding gravitational settling and mixing (\citealt{2019MNRAS.488.1332L}), to kinetic modelling (\citealt{2016A&A...594A..48L,2018A&A...615A..97L}) as part of radiation-hydrodynamics atmosphere simulations.  The data driven retrieval is also hampered by the need for a fast representation of clouds as opacity sources (e.g. \citealt{2020MNRAS.497.4183B,2020AJ....160..280C,2020ApJ...904...25L}) and it is utilised routinely for atmospheric abundance measurements (e.g. \citealt{2018Natur.557..526N,2020MNRAS.tmp.2941S}).

Modelling the formation of cloud condensation nuclei (CCNs) is key to modelling cloud formation in brown dwarf and exoplanet atmospheres because it determines when and where cloud formation sets in. Given that the formation of exoplanet CCNs is largely unexplored, \cite{2019A&A...622A.121O} introduced the nucleation rate (CCN formation rate) as a parameter into their physical model for cloud formation and transport, and they demonstrate that the nucleation rate is rather ill constrained by present observations for giant gas planets. On Earth, the CCN rate can be adjusted according to detailed observations of sand storms, bush fires, salt spray, and other factors. Nevertheless, the effect of CCN on cloud formation is the largest source of uncertainty in climate modelling on Earth, as described by the Intergovernmental Panel on Climate Change (IPCC) (\citealt{IPCCAR5SPM}). 

The James Webb Space Telescope (JWST) guaranteed-time observer (GTO) target list contains a substantial number of gas giants, including HD\,189733b, WASP-43b, and WASP-121b, {in addition to the rocky planets of the TRAPPIST-1 system, for example.} In order for cloud formation to occur in gas giants, brown dwarfs, and hot rocky planets, cloud condensation nuclei need to form from the gas phase. The quest for the primary condensate is not new and leads back to the Asymptotic Giant Branch (AGB) star dust formation modelling (\citealt{1994LNP...428..163S}). {Withing the research of AGB star mass loss, this idea has recently been re-addressed by \cite{2019MNRAS.489.4890B}.} Using the modified classical nucleation theory with surface tension values for a wide variety of materials, \cite{2018A&A...614A.126L} suggest that a considerable range of CCN materials may be available to kick-off cloud formation in exoplanet and brown dwarf atmospheres. However, little information exists about their formation mechanisms in astrophysical environments. Works by \cite{1999EPJD....6...57P,2005EPJD...32..329P},\cite{2015JPCA..119.8944L}, \cite{2017A&A...608A..55D} and \cite{gobrecht_2018} explore the formation of corundum (Al$_2$O$_3$ solid) through the study of Al-O clusters, and \cite{2019MNRAS.489.4890B} propose the Al$_2$O$_3$ solid as the most efficient condensation seed in AGB stars. \cite{2013Ap&SS.347..315C} explore the formation of  Fe-C-H ring molecules (FeC$_2$H$_{\rm n}$ and and FeC$_3$H$_{\rm n}$ for n=0,2,4) while \cite{2014CPL...612...39P} explore Ti-C clusters (Ti$_{\rm x}$C$_{\rm y}$, x, y = 1$\ldots$4). \cite{2005PhRvB..72w5402C}, considering the neutral and charged inorganic cage molecules Kr@Y$_{12}$@Z$_{20}^{\rm q}$ (Y=Ni, Pd; Z=As, Sb, Bi; q$=0,-1,-3$), report a stable system where a krypton atom is enclosed by a fullerene-like inorganic double cage. Only the exploration of the Ti-O clusters by \cite{2000JPhB...33.3417J} followed by \cite{2015A&A...575A..11L} led to the application of the cluster data to exoplanet cloud formation studies as part of kinetic cloud formation modelling by  \cite{2004A&A...414..335W},\cite{2006A&A...455..325H},\cite{2019A&A...631A..79H} and \cite{2020A&A...639A.107S}.
The most exhaustive study of the formation process of CCN has been conducted for the Earth's atmosphere, where simulations can be paired with laboratory works and in-situ measurements for the H$_2$SO$_4$-complex (\citealt{2013PhLA..377.2343S,2016JGRA..121.8152S,2017NatCo...8.2199S,2020E&SS....701142S, Dunne2016}). Recently, a 3D particle Monte Carlo (3D MC) approach has been developed to model the early stages of CCN formation for the Earth's atmosphere (\citealt{2018JCoPh.363...30K,2020AerST..54.1007K}). This model, which tracks single molecules and clusters in a 3D space, captures fluctuations that are not included in most models but are essential for the formation of stable clusters \citep{olenius2018}.

This paper focuses on the study of seed formation in exoplanet and brown dwarf atmospheres, and in particular on the formation of TiO$_2$ nucleation. TiO$_2$ has long been debated as a primary seed forming species (\citealt{1994LNP...428..163S,2013RSPTA.37110581H}) and we aim to explore the formation of TiO$_2$-CCN by two different methods and compare the results. Here we couple methods developed within astrophysics with models developed for terrestrial atmospheric science in order to study the growth of TiO$_2$ clusters. The two methods are the 3D particle Monte Carlo approach where we expand the approach presented by \cite{2018JCoPh.363...30K}, and the kinetic approach that solves a set of rate equations that require thermodynamic data for the individual clusters (\citealt{1994LNP...428..163S}). We use the Monte Carlo approach to gain more in-depth understanding about the onset of the nucleation process, the transition from a time-dependent to a stationary process, as well as the effect of reaction efficiencies. We are particularly interested in the time evolution of individual and cumulative cluster nucleation rates in 3D space.

Section~\ref{s:appr} outlines our approach and the required theoretical background for both methods applied. Section~\ref{s:MCTiO2} presents the results on the TiO$_2$-cluster formation, their individual abundances, size distributions, and the effect of the reaction efficiency $\alpha$. Section~\ref{s:comp} presents the comparison of the results from both methods, MC and rate equations approach. We discuss our results in~Section \ref{s:discussion} and conclude with Section~\ref{s:concl}.

\section{Approach}\label{s:appr}

We approach the modelling of the formation of TiO$_2$ cloud condensation nuclei (CCNs) by applying two different methods, a 3D MC approach (Sect.~\ref{ss:MC_theory}) and a steady-state approach utilizing thermodynamic cluster data (Sect.~\ref{ss:kinap}) in order to understand TiO$_2$ cluster formation as a necessary step to the formation of TiO$_2$ nucleation seeds. The MC approach enables the study of the formation dynamics of the cluster growth, which is not accessible by a steady-state approach. Each method makes assumptions and requires certain input data that we outline in the following.

\subsection{Terminology}\label{ss:Terms}
As this work combines expertise from astrophysics and Earth science, there can be a confusion due to the difference in how various terms are applied. Here we explain how we use certain terms.
A ``cluster'' is a collection of monomers sticking together through physical or chemical interaction. ``Nucleation'' is the formation of a thermodynamically stable cluster (i.e. crossing the critical size), the genesis of clusters of all other sizes is referred to as formation. ``Cloud condensation nuclei'' (CCN) are clusters large enough that they can be activated by another gas with a given supersaturation to form cloud drops (on Earth this gas is water). ``Growth efficiency'' and ``sticking probability'' is used interchangeably to indicate the probability that a collision between clusters leads to growth. ``Reaction'' refers to all processes changing the cluster size.

\subsection{Three-dimensional Monte Carlo simulation of TiO$_2$ seed formation}\label{ss:MC_theory}

We use the particle 3D MC code introduced by \cite{2018JCoPh.363...30K}. This code tracks individual particles in a 3D space and in time, and hence allows a detailed study of the nucleation process, including rare events and fluctuations for individual cluster sizes.  The computational time is large and it scales with the number of particles considered.

\begin{figure}
    \includegraphics[width=9.0cm]{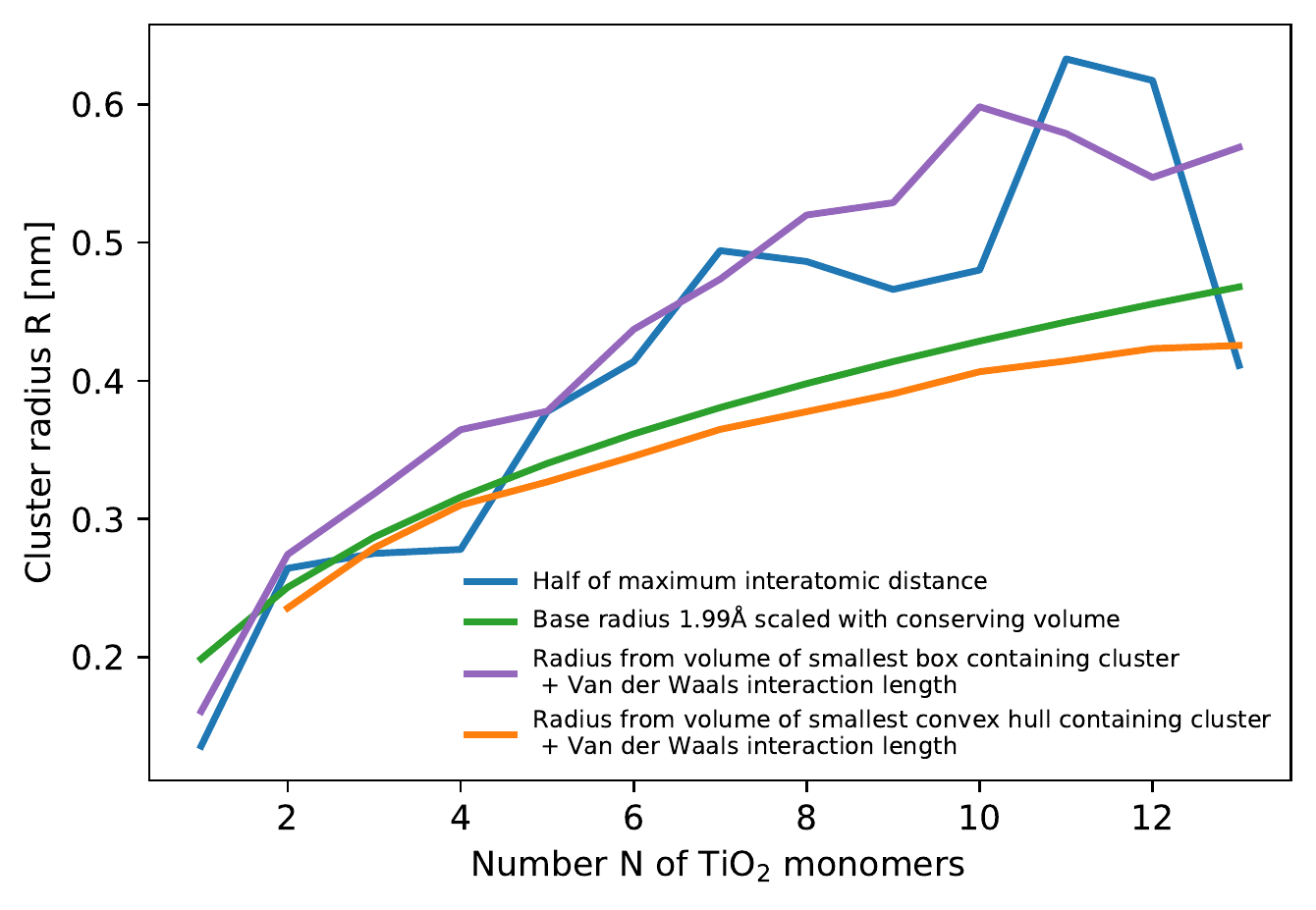}
    \includegraphics[width=9.0cm]{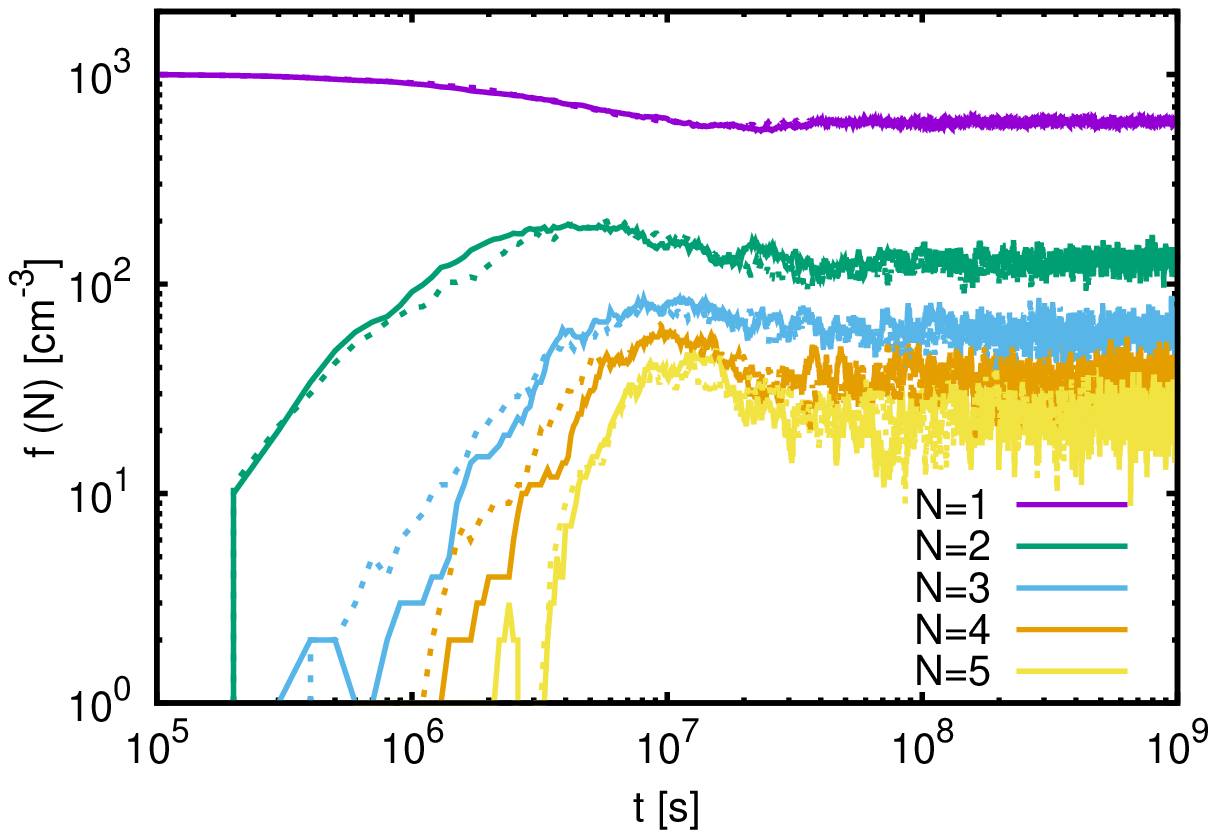}
    \caption{Top: Different cluster radius definitions. Blue line: Radius based on the maximal inter-atomic distance from DFT (density functional theory) simulations \citep{2014JCTC}. Orange line: Radius based on the ConvexHull volume approximation of cluster geometry from \citet{2014JCTC} with added Van der Waals interaction distance. Green line: $R(N)=R(1)\cdot \sqrt[3]{N}$ (Eq.~\ref{eva.4}). Purple line: Radius based on volume of smallest box containing cluster with added Van der Walls interaction distance. Bottom: Cluster number densities from MC code using different radius definitions. Solid lines: $R(N)=R(1)\cdot \sqrt[3]{N}$ (Eq.~\ref{eva.4}); dotted lines: box approximation/van-der-Waals radius (purple line in upper panel). 
    }
    \label{fig:R(N)}
\end{figure}

\noindent
Monte Carlo particles: The particles that are considered in the 3D MC simulations are characterized by size, composition, and their position in Cartesian coordinates. TiO$_2$ monomers and clusters can be traced. The largest cluster size achieved in the present 3D MC simulations is $N>10^5$ monomer units per cluster for a simulation covering $t=10^9$s. The TiO$_2$ monomer mass and radius are $m_{\textnormal{TiO}_2}\approx 1.326206\cdot 10^{-25}$ kg and $r_{\textnormal{TiO}_2}\approx 0.199$ nm, respectively. The definition of cluster radii for sizes larger than the monomer ($N>1$) is challenging, and we note that larger clusters can be more compact than smaller clusters with less elongated structures. This is explored in Fig.~\ref{fig:R(N)} where the results from different methods to calculate a cluster diameter are shown. To allow for comparison, the interacting volume of each cluster is calculated with the respective method and the radius of a sphere with the equivalent volume is plotted. The interacting volume of each cluster is larger than the geometrical volume assuming spherical particles because of van der Waals forces determining the actual chemical interaction. The interacting volume  is calculated by finding the axes of the smallest box that contains the cluster of size $N$ and adding the Van der Waals interaction length of $1.6\AA$ \citep{Koch2017} to each of the axes. The volume of that extended box is the interacting volume of the cluster. For a different method, plotted in orange, the volume was calculated using Delaunay triangulation and the ConvexHull algorithm as described by \cite{Barber1996}. This algorithm uses polygons to approximate the surface of the cluster and therefore gives a smaller volume than the approximation through a box. After determining the sphere equivalent radius from this volume, the Van der Waals interaction length is added to account for the interaction distance that is larger than the physical dimensions spanned by the atomic cores of the cluster. In our 3D MC simulations, the cluster collisions are treated as a hit-and-stick coagulation process between clusters, hence contractions, compactification, or expansion are omitted.

\begin{figure}[h]
    \centering
    \includegraphics[width=9.5cm]{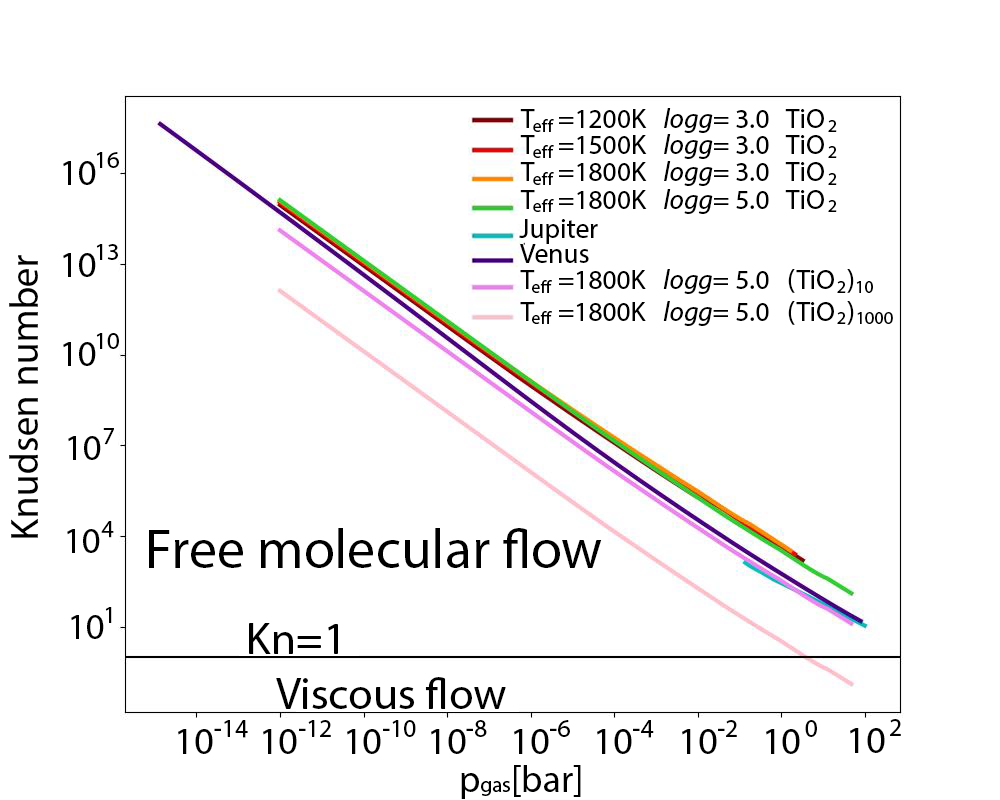}
\caption{Knudsen number, $Kn = l_{\rm MFP}/(2a)$ ($l_{\rm MFP}$ -- mean free path, $a$ -- particle radius) for different exoplanet and brown dwarf {\sc Drift}-Phoenix model atmosphere results, except for Jupiter and Venus. The number $Kn$ indicates whether the cluster interactions occur through diffusion ($Kn <1$) or in a free molecular flow ($Kn >1$). Exoplanet and brown dwarf atmospheres are predominately in the regimes of a free molecular flow ($Kn >1$). {Knudsen numbers for Earth's atmosphere and typical cloud particle sizes are $\approx 10^{-5}$and therefore fall in the turbulent, viscous case.}}
    \label{fig:hKn}
\end{figure}

\noindent
Monte Carlo cluster growth: The frequency with which cluster collisions may occur depends on the ambient density. Figure~\ref{fig:hKn} demonstrates that exoplanet atmospheres that fall into the thermodynamic regime of giant gas planets shown here are mainly in the hydrodynamic regime of a free molecular flow.  Hence, the collision frequencies of the clusters in our 3D MC model will be determined by their mean free path. We note that diffusion is commonly assumed to determine the collision frequencies in molecular-dynamic simulations for Earth atmosphere nucleation studies (see e.g. \citealt{2018JCoPh.363...30K,2020AerST..54.1007K}), which is supported by a Knudsen number of $10^{-15}$ for Earth's atmosphere (and hence as an example for rocky planets).

The calculation of the time interval, $\Delta t$,  between two possibly constructive cluster collisions is therefore guided by an analysis of the Knudsen number (Fig.~\ref{fig:hKn}). The Knudsen number ($Kn = l_{\rm MFP}/(2a)$; $l_{\rm MFP}$ -- mean free path, $a$ -- particle radius) is derived for results from 
{\sc Drift}-Phoenix model atmosphere simulations (\citealt{2009A&A...506.1367W, 2011A&A...529A..44W}). Results include the hydrostatic $(T_{\rm gas}, p_{\rm gas})$-structures, which are  calculated consistently with cloud formation, chemical equilibrium, and gas and cloud radiative transfer. As a result, the cloud structure can be described by mean particle size, $a$, material volume fraction, and element depletion. The maximum MC time step is therefore defined as
\begin{equation}
\max(\Delta t) = \frac{l_{\rm MFP}}{v_{\rm rel}},
\end{equation}
with the local thermal velocity $v_{\rm rel}(T)$, and the mean free path $l_{\rm MFP}$ within the gas (see e.g. Eq. (10) in \citealt{2003A&A...399..297W}). Hence for temperatures between 500 K and 2000 K, max($\Delta t$) lies on the order of $10^7$ s; for better numerical resolution, we have chosen $\Delta t=100$ s.
The clusters move with their local thermal velocity for a time $\Delta t$. If their location overlaps after $\Delta t$, they are considered to merge. The merging can occur between clusters of different sizes and a monomer only (polymer-monomer) or between all available sizes (polymer-polymer).
Such a merging of two MC particles depends on their collision probability but also on the stickiness (i.e. reactiveness) of the two interacting clusters. This is parameterised as a sticking probability, $\alpha$. The effect of the sticking probability on the cluster growth is tested by simulating different sticking probabilities $\alpha\in[0,1]$ and checking for $r\le\alpha$ with a random number $r\in [0,1)$ once two particles overlap. Only if this condition is fulfilled are two MC particles  merged (coagulated in Earth science terms) by adding their volumes and their masses. After coagulation, a new monomer is added if the total number of individual particles is less than $10^3$ such that $df_{\rm tot}/dt\sim 0$ for the total particle density $f_{\rm tot}$. By adding monomers to the simulation domain, we simulate a constant influx of monomers whilst we keep the total particle density constant. As such, we take into account that cloud formation requires the transport, diffusion, and mixing of particles.

Besides the short-term interaction due to the collision of TiO$_2$ clusters, long-term interactions are caused by the fact that TiO$_2$ clusters are dipoles. TiO$_2$ has a dipole moment of approximately $7D$ \citep{bruenken_2008} where $D\approx 3.33\cdot 10^{-30}$ C$\cdot$m is the debye unit of the electric dipole moment.  Small TiO$_2$ clusters show dipole moments with similar values to the TiO$_2$ monomer (e.g. (TiO$_2$)$_3$ and (TiO$_2$)$_4$), or no dipole moment like the symmetric dimer (TiO$_2$)$_2$. However, the dipole contribution to the velocity is on the order of $v_{dip}\sim 2p \Delta t/(r^3\cdot m)\approx 10^{-26}$ m s$^{-1}$ for $r=l_{\rm MFP},\ \Delta t=10^7$ s and $m=m_{\textnormal{TiO}_2}$ , which is much smaller than the thermal velocity that is on the order of several hundreds of m s$^{-1}$; hence the dipole contribution is negligible.

\begin{figure}
    \includegraphics[width=8.6cm]{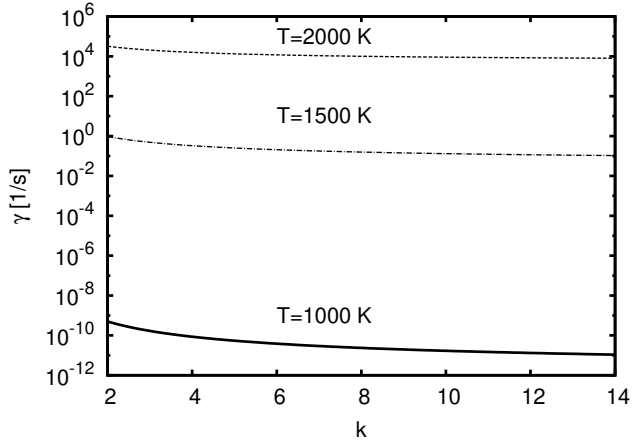}
    \caption{Evaporation frequency $\gamma(N)$ (Eq.~\ref{eq:eva.1}) for TiO$_2$ clusters as a function of their size $N$ for different gas temperatures.}
    \label{fig:eva}
\end{figure}

\smallskip
\noindent
Monte Carlo cluster evaporation: Clusters not only undergo constructive growth processes but can also evaporate if they are thermally unstable. The most unstable cluster is called the critical cluster, $N_*$, and all clusters larger than $N_*$ will preferably grow and therefore be thermally stable. After every time step $\Delta t$, we check the probability of evaporation, $\gamma\cdot\Delta t$, for particles that consist of more than one monomer. If  $r\le 1-\exp\left(-\gamma\cdot\Delta t\right)$ for a random number $r\in[0,1)$, evaporation is simulated by emitting a single monomer randomly. We note that clusters are not arranged as a collection of smaller clusters but as a collection of monomers; hence the probability of a polymer escaping is quite small and we only model the emission of single monomers from large clusters.

The evaporation frequency, $\gamma$, for clusters of size $R(N)$ and mass $m(N)$ is described according to \cite{yu_2005} as
\begin {eqnarray}
\gamma&=&\sqrt{\frac{8\pi k_B T(m(1)+m(N))}{m(1)m(N)}}\left(R(1)+R(N)\right)^2 n_{a,sol}^{\infty}(T) \times \nonumber \\
&\times& \exp\left(\frac{2M_1 \sigma(T)}{\varrho(N)\cdot \mathcal{R}\cdot T R(N)}\right) \label{eq:eva.1}
,\end {eqnarray}
where $n_{a,sol}^{\infty}$ is the phase-equilibrium vapour density for TiO$_2$ with respect to the solid TiO$_2$[s] phase \citep{woitke_2018}, the surface tension $\sigma$ \citep{2015A&A...575A..11L}, and the density $\varrho(N,T)$  of the clusters of size $N$ are
\begin {eqnarray}
n_{a,sol}^{\infty} [\textnormal{m}^{-3}] &=& \frac{10^{-1}}{k_B T[\textnormal{K}]}\times \nonumber \\
&\times&\exp\left[-\frac{7.70443\cdot 10^4}{T[\textnormal{K}]}+40.3144 \right. \nonumber\\
&-& 2.59140\cdot 10^{-3}\cdot T[\textnormal{K}]+6.02422\cdot 10^{-7} \cdot T[\textnormal{K}]^2 \nonumber \\[0.2cm]
&-& \left. 6.86899\cdot 10^{-11}\cdot T[\textnormal{K}]^3\right] \label {eva.2}\\
\sigma [\textnormal {N m}^{-1}] &=& (535.124-0.04396\cdot T[\textnormal{K}])\cdot 10^{-3} \label {eva.3} \\
\varrho(N) &=&\frac{4m(N)}{4\pi R(N)^3}, \label{eva.4}
\quad\quad R(N)= R(1) \cdot \sqrt[3]{N}, 
\end {eqnarray}
with a monomer radius of $R(1)=r_{\textnormal{TiO}_2}\approx 0.199$ nm and a monomer mass of $m(1)=m_{\textnormal{TiO}_2}\approx 1.326\cdot 10^{-25}$ kg. Boltzmann's constant is $k_B\approx 1.38\cdot 10^{-23}$ J K$^{-1}$ , $T$ is the temperature, $M_1\approx 80$ g mol$^{-1}$ is the molar mass of TiO$_2$ , and $\mathcal{R}\approx 8.31$ J (mol k)$^{-1}$ is the universal gas constant. Thus the formation of clusters is determined by the relation between the collision and evaporation frequencies, which are both increasing functions of temperature. Using the evaporation rate (\ref{eq:eva.1}) is the standard approach to include evaporation \citep{seinfeld_2006,yu_2005} as experimental evaporation rates are hard to obtain. For H$_2$SO$_4$-H$_2$O clusters, we have validated this approach against experimental results \citep{2018JCoPh.363...30K}.

Figure~\ref{fig:R(N)} (top panel) compares the values used in the 3D MC simulation (green line, Eq.~\ref{eva.4}) to other representations of cluster diameters. 
Numerical checks (Fig.~\ref{fig:R(N)}, lower panel) of the effect of these cluster radius differences showed little effect of these uncertainties on the results of the present study; hence the assumption of spherical particles works sufficiently well although small particles deviate from a spherical form. 

By applying the surface tension we assume our clusters to have a non-planar surface. The Kelvin effect (i.e. the curvature) is taken into account through the exponential term in Eq.~(\ref{eq:eva.1}) (Kelvin equation), which models the effect of the cluster surface curvature. The same effect is accounted for in the rate equation approach (Sect.~\ref{ss:kinap}) when the surface energy of cluster $N$ is expressed as Eq. (4.14) in \citep{2013RSPTA.37110581H}. 

Figure \ref{fig:eva} shows the evaporation frequency (Eq.~\ref{eq:eva.1}) for TiO$_2$ clusters as a function of constituting monomers $N$ for different temperatures. 
The strong temperature dependence suggests that evaporation will be negligible for $T\le 1000$ K whilst evaporation becomes dominant for $T\ge 2000$ K.

\smallskip
\noindent
Monte Carlo cluster formation rates: The time evolution of the formation rate of clusters larger than size $N$ is derived from our 3D MC results as
\begin {eqnarray}
\label{eq:MCJ*}
J_{\rm N}(t)=\frac{f_{\ge N}(t)}{t}
,\end {eqnarray}
where $f_{\ge N}(t)$ is the number density of clusters above a given size $N$ as a function of time $t$ within the volume $V$ ($1$ cm$^{-3}$  for the present simulations). Thus, the number of clusters of size $N$ is  $f_{=N}(t) \cdot V$ (see Fig.~\ref{fig:size_distr_1}).

\subsection{Kinetic seed formation through rate equations}\label{ss:kinap}

The kinetic, steady-state approach describes the temporal evolution of the size distribution function as \citep{2013RSPTA.37110581H}
\begin{equation}
\label{eq:master}
\frac{df(N,t)}{dt} = \sum_{i=1}^I J^c_i(N,t) -  \sum_{i=1}^I J^c_i(N+i,t),
\end{equation}
where $f(N)$ is the number density of a molecular cluster containing $N$ $i$-mers 
contributing to the growth of the
particle and $J^c_i(N,t)$ is the effective flux (or transition rate) for
the growth of the particle of size $N-i$ to size $N$ due to all
associating or dissociating reactions involving an $i$-mer of the
condensing species. This stationary flux through cluster space is 
\begin{equation}
\label{eq:clflux}
J^c_i(N,t) = \sum^{\rm R_i}_{\rm r_i -1} 
            \left(\frac{f(N-i, t)}{\tau_{\rm gr}(r_i, N-i, t)} - \frac{f(N, t)}{\tau_{\rm ev}(r_i, N, t)}\right), 
\end{equation}
summing over all chemical reactions $r_i$ in
which an $i$-mer is involved.  Equation~\ref{eq:clflux}  is equivalent to Eq. (15) in \cite{2018JCoPh.363...30K}, both giving the number of seed particles cm$^{-3}$ s$^{-1}$.
The growth time $\tau_{\rm gr}(r_i,
N-i, t)$ is the time by reaction $r_i$ leading
from cluster size $N-i$ to cluster size $N$. The evaporation time $\tau_{\rm ev}(r_i, N, t)$ is the time leading from size $N$ to size $N-i$,
\begin{eqnarray}
\label{eq:taugr}
\frac{1}{\tau_{\rm gr}(r_i, N-i, t)}\!\!\!\!\!&=&\!\!\!\!\!A(N-i)\, \alpha(r_i, N-i) v_{\rm rel}(n_{\rm f}(r_i), N-i)\, n_{\rm f}(r_i)\\
\label{eq:tauev}
\frac{1}{\tau_{\rm ev}(r_i, N, t)}\!\!\!\!\!&=&\!\!\!\!\!A(N)\, \beta(r_i, N)\, v_{\rm rel}(n_{\rm r}(r_i), N)\, n_{\rm r}(r_i).
\end{eqnarray}
The densities $n_{\rm f}(r_i)$ and $n_{\rm r}(r_i)$ are the number density of the molecule of the growth (forward) process and of the evaporation (reverse) process for reaction $r_{\rm i}$,
respectively \citep{1998A&A...337..847P}. The surface $A(N)$ is the surface of the cluster of size $N$. The average relative velocity between the
growing or evaporating TiO$_2$ molecule and the cluster is 
$v_{\rm rel}$, and is defined as the thermal velocity 
\begin{equation}
v_{\rm rel} = \sqrt{ \frac{k_B T}{2\pi} \left(\frac{1}{m_{\rm N}} + \frac{1}{m_{\rm TiO_2}}\right)}.
\end{equation}
 The reaction efficiencies for growth and evaporation via reaction $r_{\rm
  i}$ are given through $\alpha(r_i, N-i)$ and $\beta(r_i, N)$. We note that $\alpha$ can also be referred to as the sticking probability. Equation~\ref{eq:tauev} corresponds to Eq.~\ref{eq:eva.1} in the 3D MC approach, except that the thermal stability does not enter Eq.~\ref{eq:tauev}. Instead,  we use the thermodynamic properties of the clusters according to  Eq.~\ref{eq:lma}.
The challenge is that the growth and evaporation efficiency
coefficients, $\alpha(r_i, N)$ and $\beta(r_i, N)$, are often unknown for the different cluster sizes $N,$ which is why we perform 3D MC simulations in addition to the kinetic approach. As Eq. (\ref{eq:tauev}) corresponds to Eq. (\ref{eq:eva.1}), hence $\gamma\sim \tau_{\rm ev}^{-1}$, we can estimate $\beta=\gamma/(A(N) v_{\rm rel}(n_{\rm r}(r_i), N)\, n_{\rm r}(r_i) )$. For $T=1000$K, $\beta$ is on the order of $\approx 10^{-1}-10^0$ for all considered sizes. Hence, we primarily use $\beta=1$ when comparing the MC approach and the kinetic model.

For further considerations, we need to introduce a reference  equilibrium state.
In that, we follow Patzer et al. (1998) who show in their Appendix A that if the temperatures of all components are equal, the supersaturation ratio of a cluster of size $N$ with respect to the bulk is $S_{\rm N}= (S_1)^{\rm N}$. Hence, phase equilibrium between monomers and between
the clusters and the bulk solid, plus simultaneous chemical
equilibrium in the gas phase, plus thermal equilibrium (i.e. all components have the same temperature) characterise this equilibrium state. In such local thermodynamic equilibrium (LTE) between the gas phase and the clusters, where $\fref(N)$ is the equilibrium number density, the principle of detailed balance holds for a single microscopic growth process and its respective reverse, that is, the
evaporation process. This implies that under the condition of
detailed balance, $\fref(N-1)/\tau_{\rm gr}(N-1)=\fref(N)/\tau_{\rm
ev}(N)$, which allows us to express the evaporation rate by the growth
rate. The
equilibrium number densities for the clusters 
and the monomers are
$\fref(N-i)$, $\fref(N)$, $\nfref(r_i)$, and $\nrref(r_i)$ .
The law of mass action links these
equilibrium particle densities to Gibbs free energies,  
\begin{equation}
\label{eq:lma}
\left( \frac{\fref(N-i)\, \nfref(r_i)}{\fref(N)\,\nrref(r_i)}\right) = 
 \exp \left(
      \frac{\dG(r_i, N, T_{\rm d}(N))}
           {R\,T_d(N)}
      \right).
\end{equation}
The energy $\dG(r_i, N, T_{\rm d}(N))$ [kJ mol$^{-1}$] is the Gibbs free energy
of formation and it can be calculated from the standard molar Gibbs free
energy of formation of all reaction participants at the temperature
$T_{\rm d}(N)=T_{\rm gas}$ , which is
\begin{eqnarray}
\label{eq:dG1}
\nonumber
\dG(r_i, N, T_{\rm d}(N)) &=& \dG(N,  T_{\rm d}(N)) - \dG(N-i,  T_{\rm d}(N))\\
                          &+& \dG(n_{\rm r}(r_i), T_{\rm d}(N))\\
                          \nonumber
                          &-&  \dG(n_{\rm f}(r_i), T_{\rm d}(N)).
\end{eqnarray}
Assuming LTE, the equilibrium cluster size distribution is therefore expressed by a Boltzmann
distribution,
\begin{equation} 
\label{eq:fref}
\fref(N) = \fref(1)\,\exp\left(-\frac{\Delta G(N)}{RT}\right),
\end{equation} 
with $\fref(1)$ being the equilibrium density of the monomer (here TiO$_2$). 

The difference in energy $\Delta G(N)$ is the free energy change due to the formation of a
cluster of size N from the saturated vapour. It is related to the
standard molar Gibbs free energy of formation of the N-cluster
$\dG(N)$ by
\begin{equation}
\label{eq:dG}
\Delta G(N) = \dG(N, T) + RT \ln\left(\frac{p_{\rm sat}(T)}{\pst}\right) - N\,\dG_1(s, T).
\end{equation}
The energy $\dG_1(s),T$ is the standard molar Gibbs free energy of the formation of the
solid phase, and $\pst$ is the pressure of the standard state. Most
often, $\pst$ is the atmospheric pressure on the Earth at which
$p_{\rm sat}$ and $\dG(N, T)$ were measured. The right hand side of
Eq.~\ref{eq:dG} now contains  quantities that can be determined from
lab experiments or quantum-chemical calculations.

No assumptions were necessary with respect to size-independent surface tensions as required in classical nucleation theory, which is the droplet model applied in Sect.~\ref{ss:MC_theory}. We note that only a classical nucleation approach (droplet model) requires the consideration of the so-called Kelvin effect. An exponential (Eq.~\ref{eq:eva.1}, also Eq. 18 in \citep{2018ApJ...860...18P}) corrects the saturation vapour pressure, $p_{\rm vap}$,  data that is derived for flat surfaces. The Kelvin effect is therefore used to include some representation of the curvature of the clusters for the evaporation process. The use of thermodynamic cluster data in combination with the assumption of detailed balance conveniently side-steps calculating evaporation such that the concept of  surface tension is not required to describe the effect of surface curvature on thermal stability.

\section{Three-dimensional Monte Carlo dynamics: (TiO$_2$)$_{\rm N}$ cluster formation results}\label{s:MCTiO2}

The results from the 3D MC simulations for TiO$_2$ cluster formation are presented for a set of local gas temperatures of $T=$500K, 1000K, 1500K, and 2000 K and a local total gas density of $f_{\rm tot, gas}=10^3$ cm$^{-3}$ of these TiO$_2$ clusters. 
These thermodynamic values are  representative of the region where cloud particle seed formation may occur in exoplanet and brown dwarf atmospheres (e.g. \citealt{2015A&A...575A..11L,2017A&A...603A.123H}). The 3D MC simulations are run for a fixed volume of $V=1$ cm$^3$ as the computational domain in which the total number of particles remains constant for each run, namely $10^3$ particles per cm$^3$. By assuring $f_{\rm tot, gas}=\,$const, each simulation is run under isobaric conditions such that $T=\,$const. Our studies are therefore not affected by changing thermodynamic conditions and all 3D MC results presented here are caused by particle interaction alone.

For each temperature, we explore two cases for cluster growth:
a) cluster-monomer growth (i.e. polymer-monomer), and
b) cluster-cluster growth (i.e. polymer-polymer) where cluster growth can proceed by inter-cluster coagulations of all sizes. The growth tree is visualised in  Fig.~\ref{fig:dectre} to help understand the presented results. In order to address the effect of parameter uncertainty,  different growth efficiencies (sticking probabilities)  $\alpha=1.0, 0.1 $ and a random value $rand$ were tested.



\begin{figure}
\includegraphics[width=8.6cm]{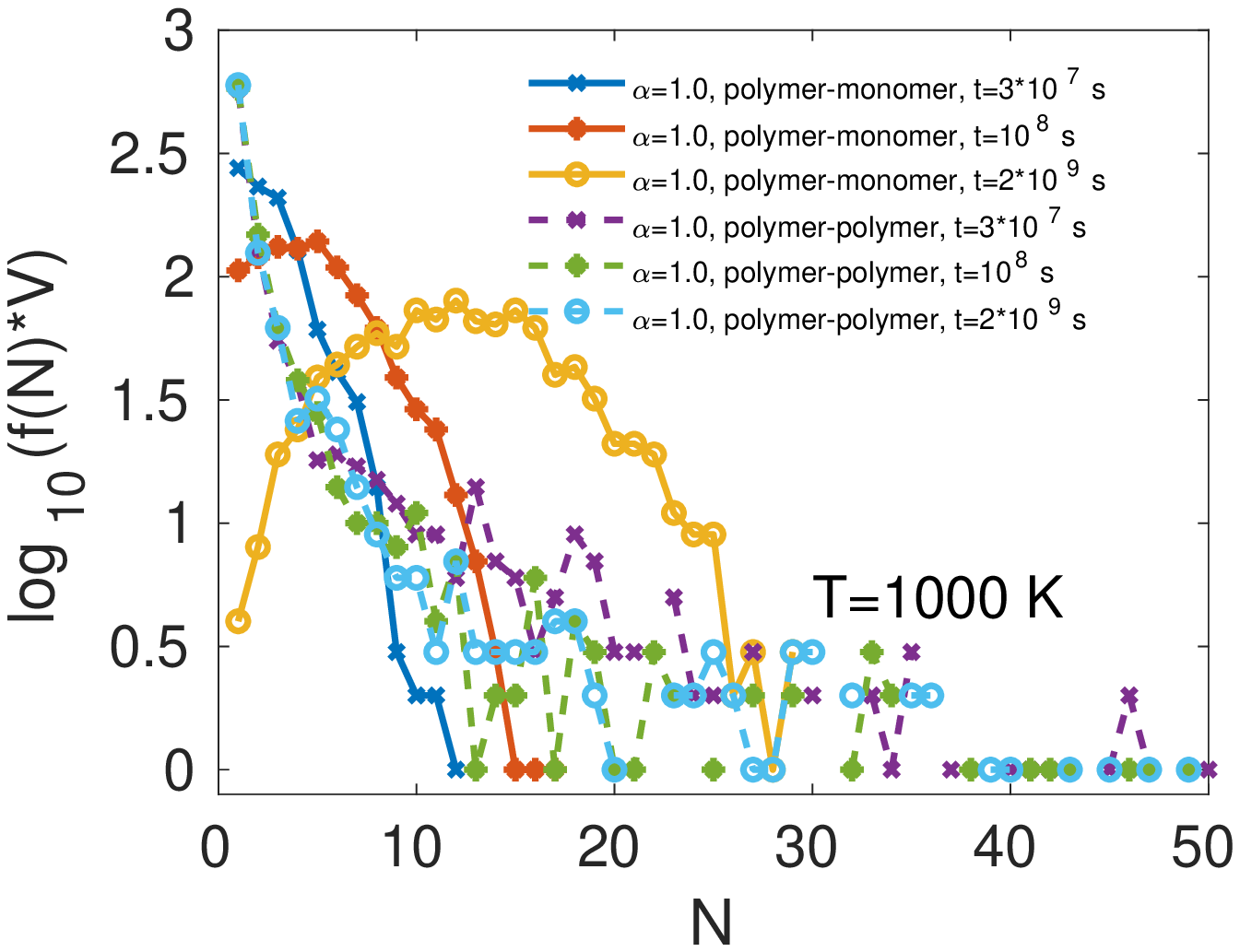}
\includegraphics[width=8.6cm]{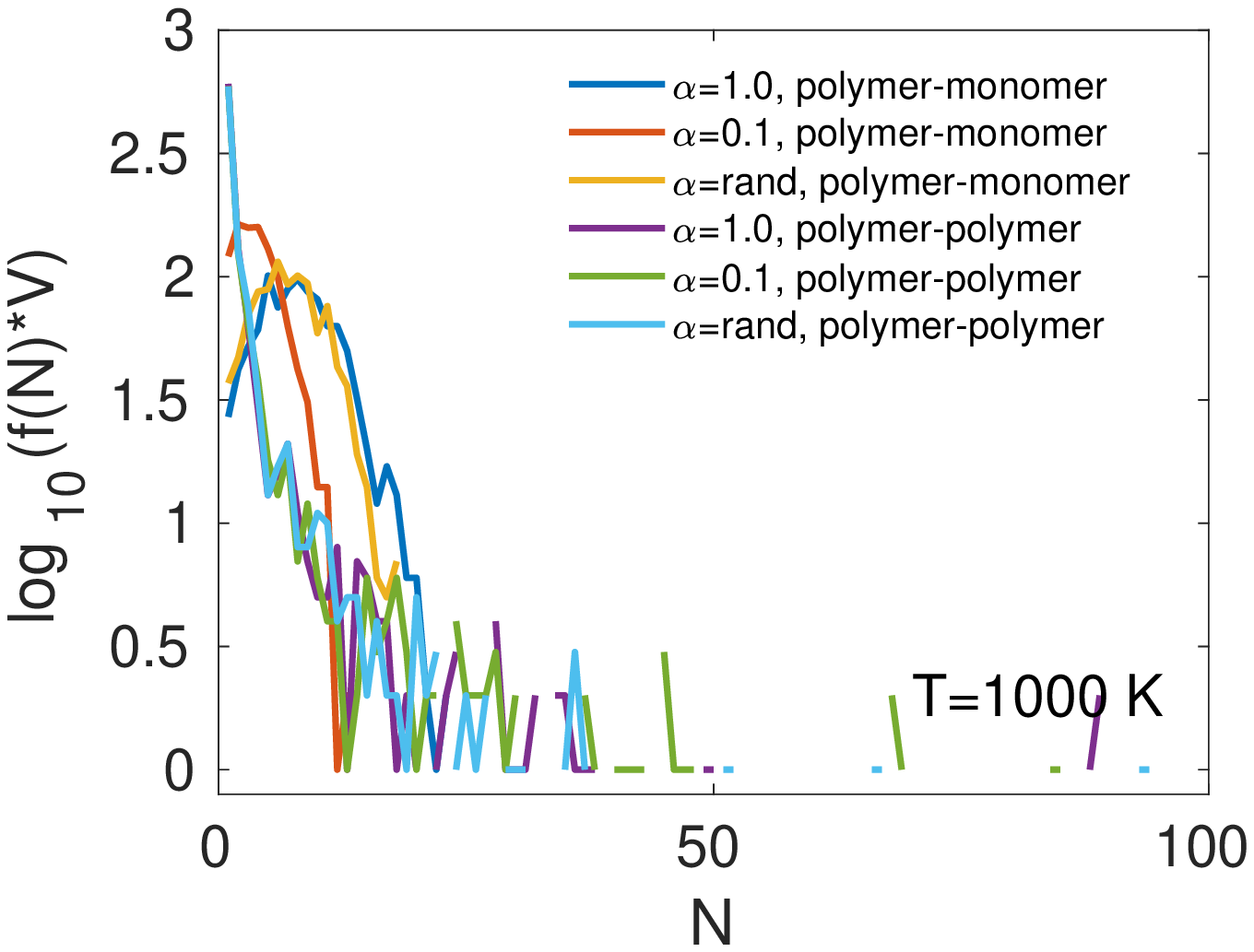}
\includegraphics[width=8.6cm]{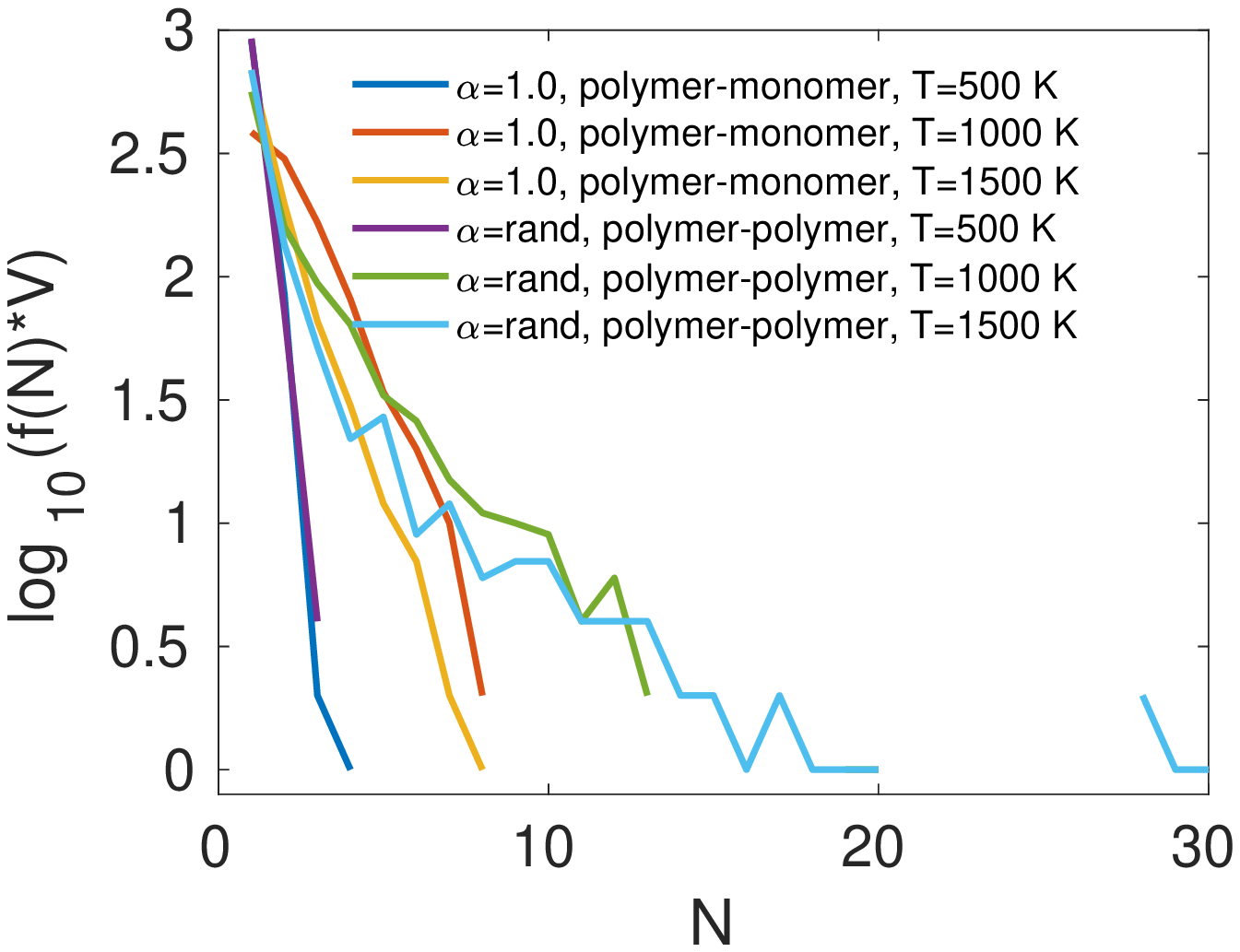} 
\caption{TiO$_2$  cluster size distribution in a homogeneous TiO$_2$ bath gas under exoplanet atmosphere conditions at $f_{\rm tot, gas}=10^3$ cm$^{-3}$.\\
{\bf Top:} Cluster size distributions  for $T=1000$K at different times for a constant $\alpha=1.0$ for poly-mono and  poly-poly.
{\bf Middle:}
Cluster size distributions for $T=1000$K and various $\alpha$ at  $t=4\cdot 10^8$s  ($rand$ - random choice of the sticking probability).
{\bf Bottom:} Comparison of size distributions for different temperatures after 
$\approx 1.7\cdot 10^7$ s.}
\label{fig:size_distr_1}
\end {figure}

\begin{figure}[h!]
    \centering
    \includegraphics[width=9.3cm]{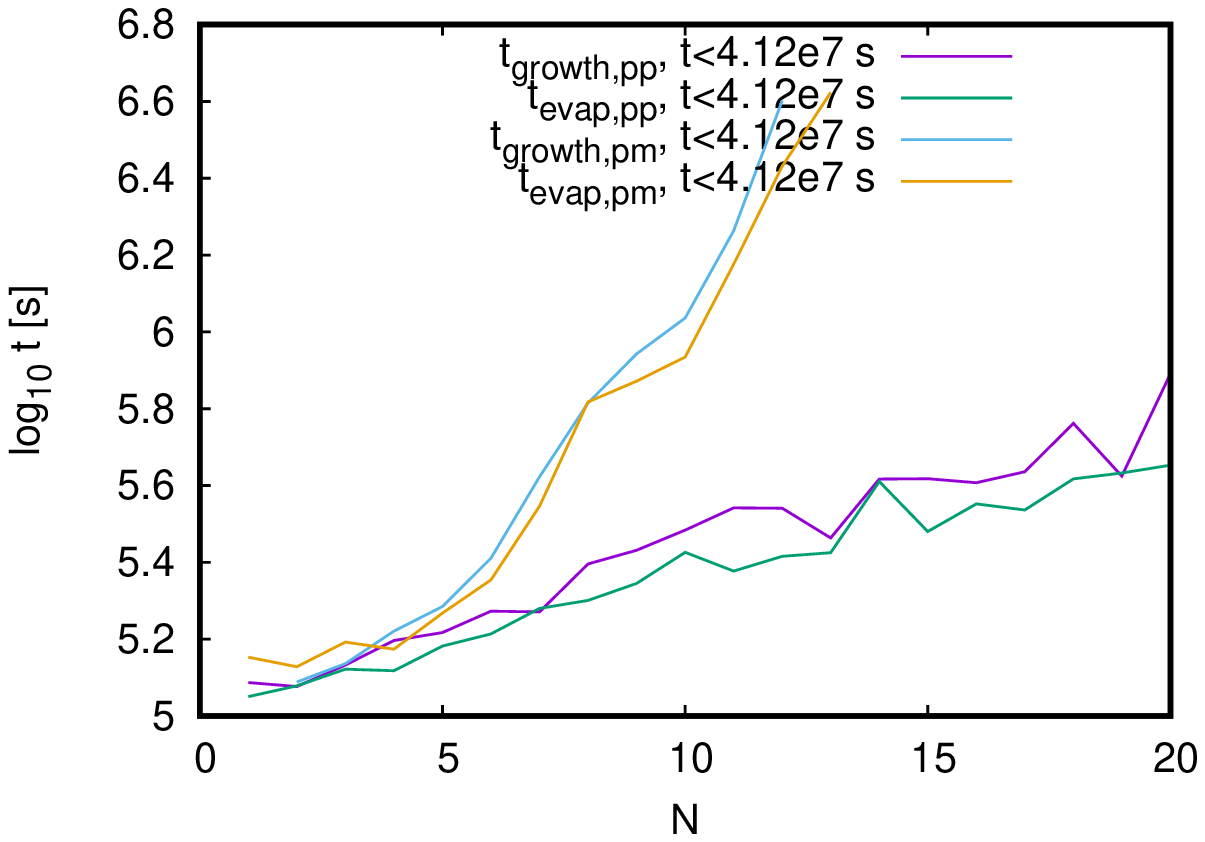}
    \includegraphics[width=9.3cm]{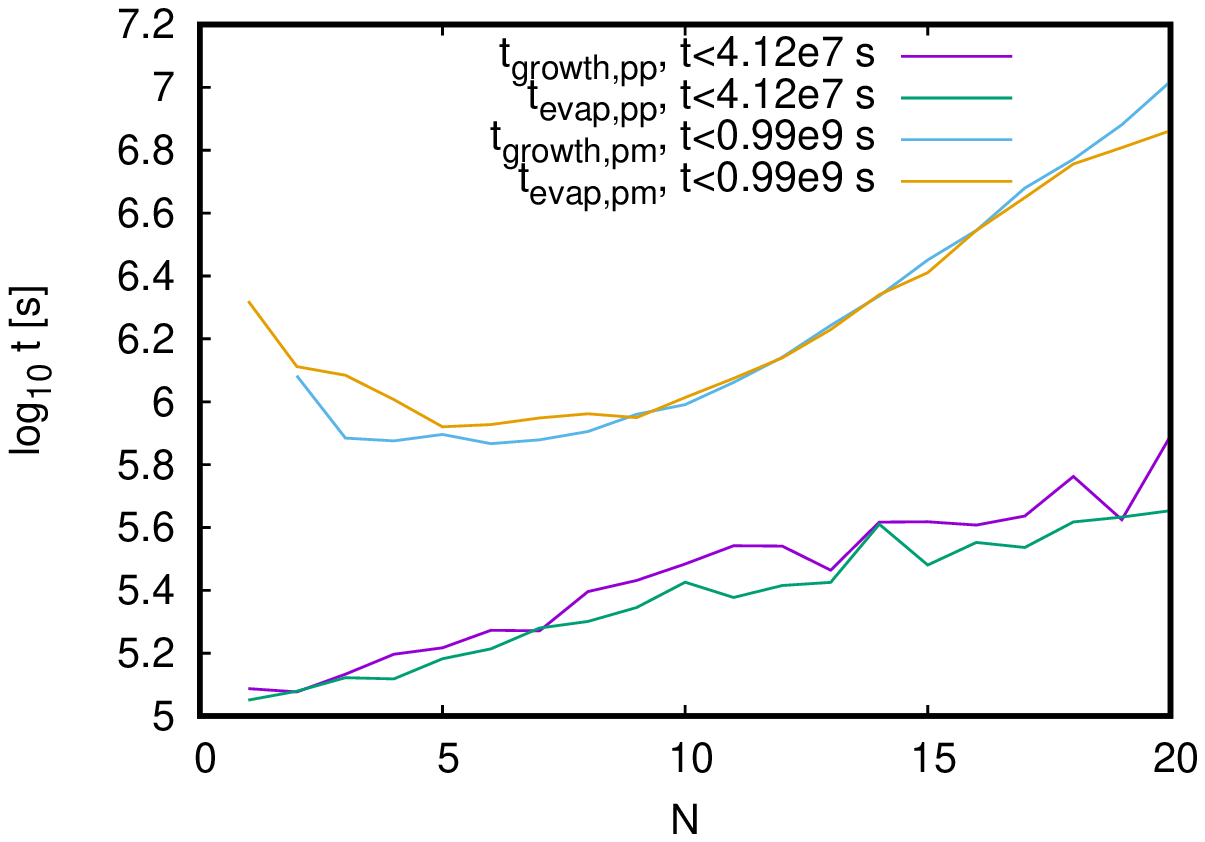}
\caption{Average (net) TiO$_2$ cluster growth and evaporation times for $\alpha=1$, $T=1000$ K, and $f_{\rm tot, gas}=10^3$ cm$^{-3}$ . pp: polymer-polymer; pm: polymer-monomer. {\bf Top:} The values are evaluated until $t_{\rm max}=4.12\cdot 10^7$ s, which corresponds to the $J_{N\ge 20}$ peak time for polymer-polymer cluster growth (see Fig.~\ref{fig:nucl_rate_1} bottom right). {\bf Bottom:} The values are evaluated until different total run times $t$ to capture the polymer-monomer peaks at $t=0.99\cdot 10^9$ s (see Fig.~\ref{fig:nucl_rate_1} bottom left). 
}  \label{fig:gr_ev1}
\end{figure}

\begin {figure*}
\includegraphics [width=8.6cm] {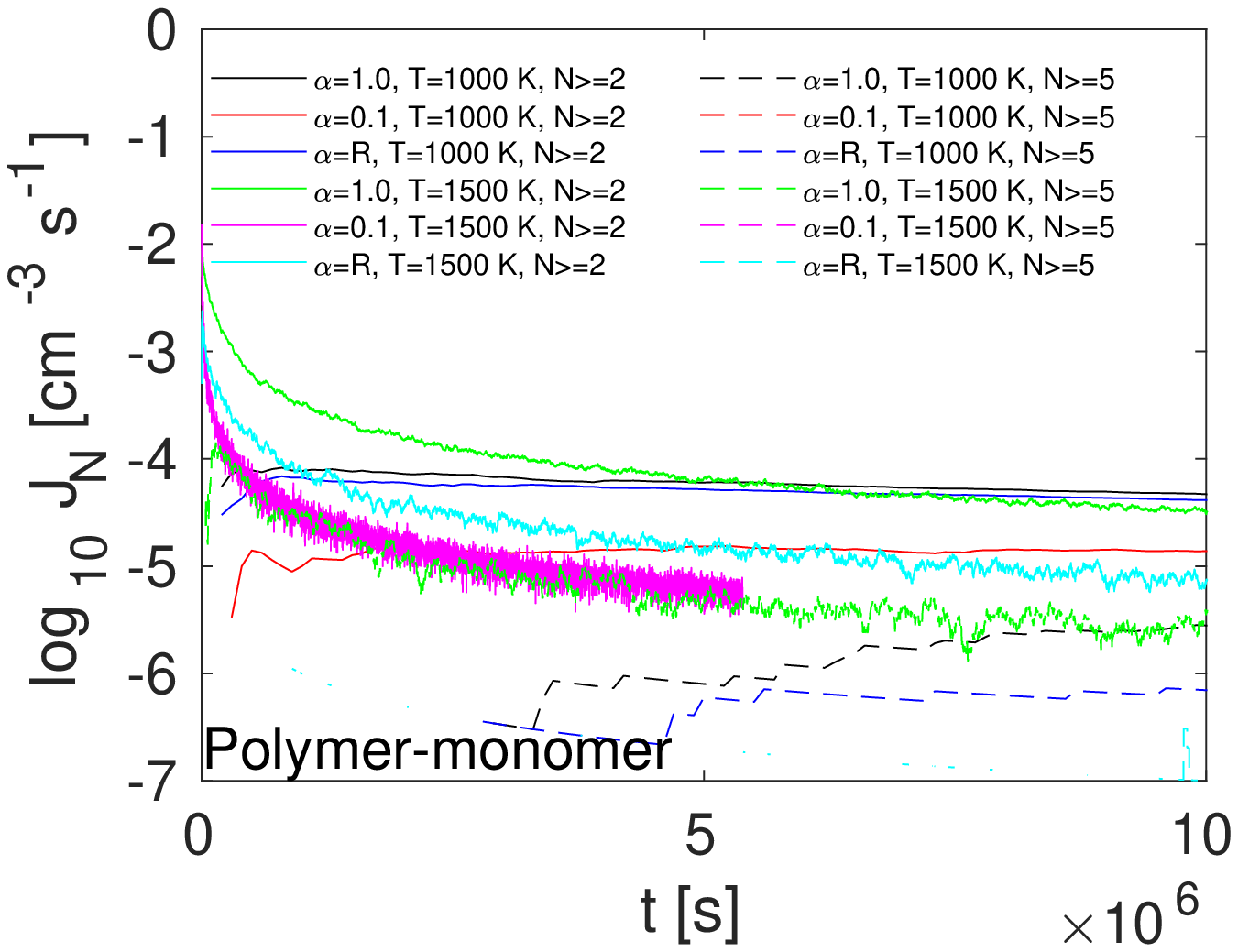}
\includegraphics [width=8.6cm] {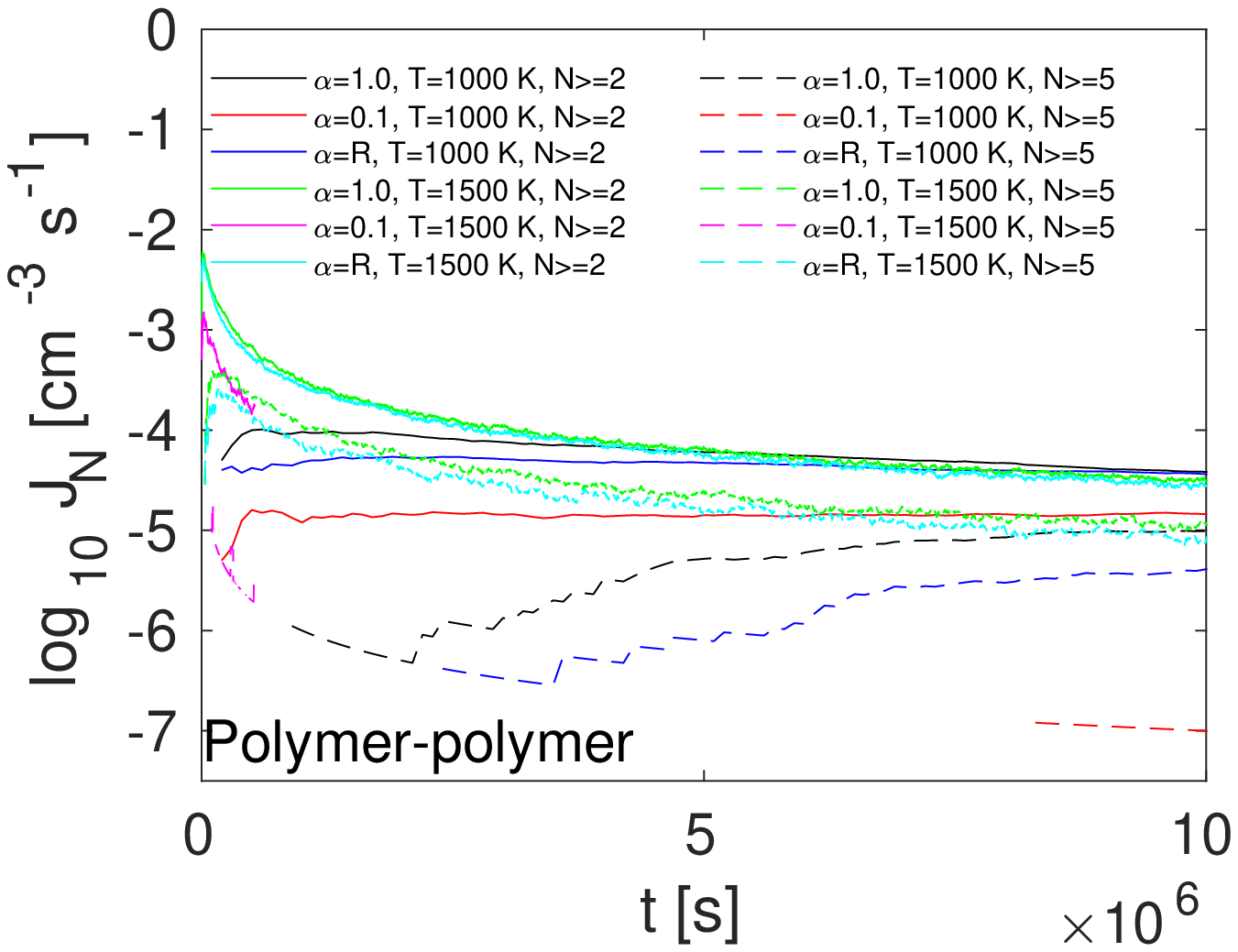}\\
\includegraphics [width=8.6cm]  {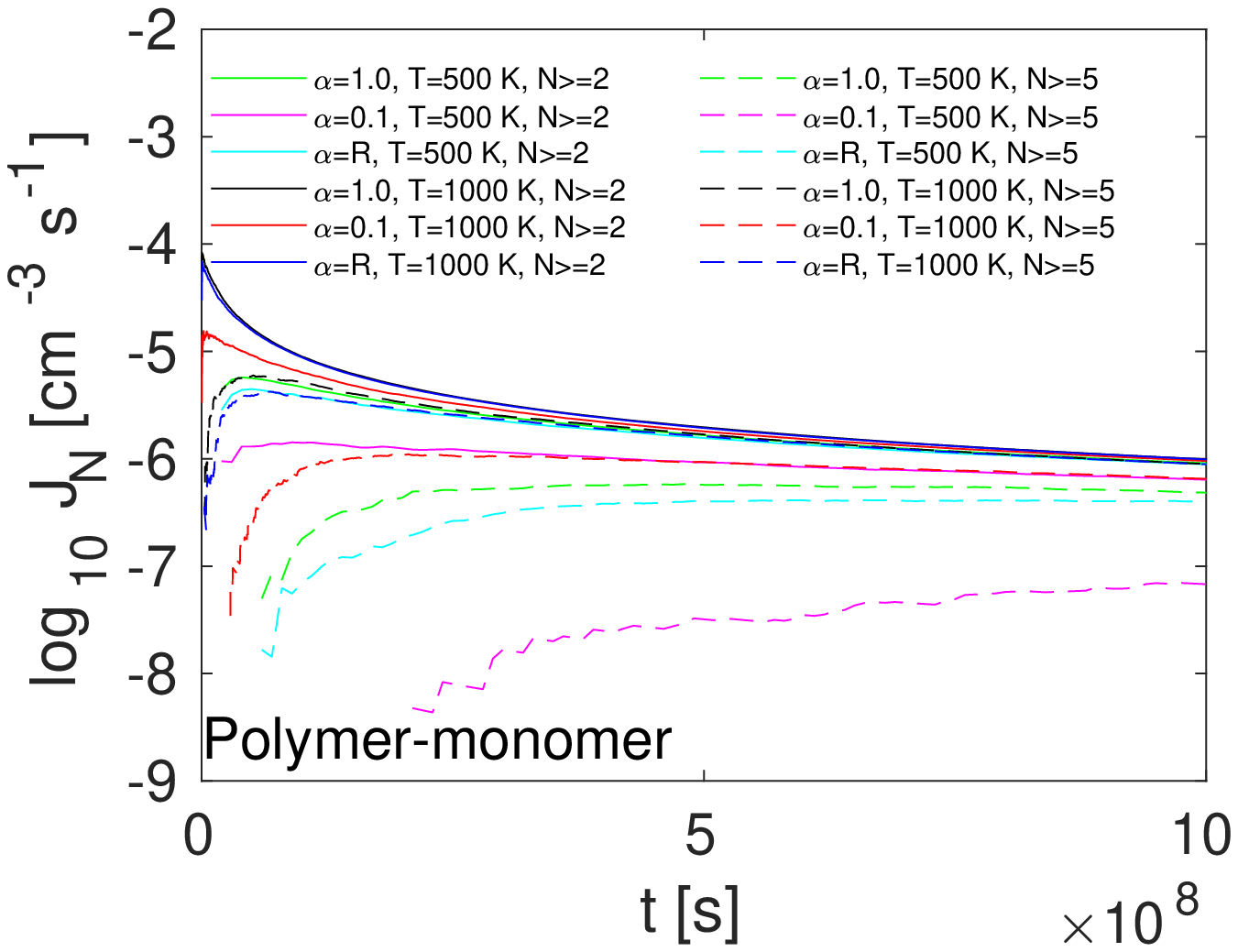} 
\includegraphics [width=8.6cm] {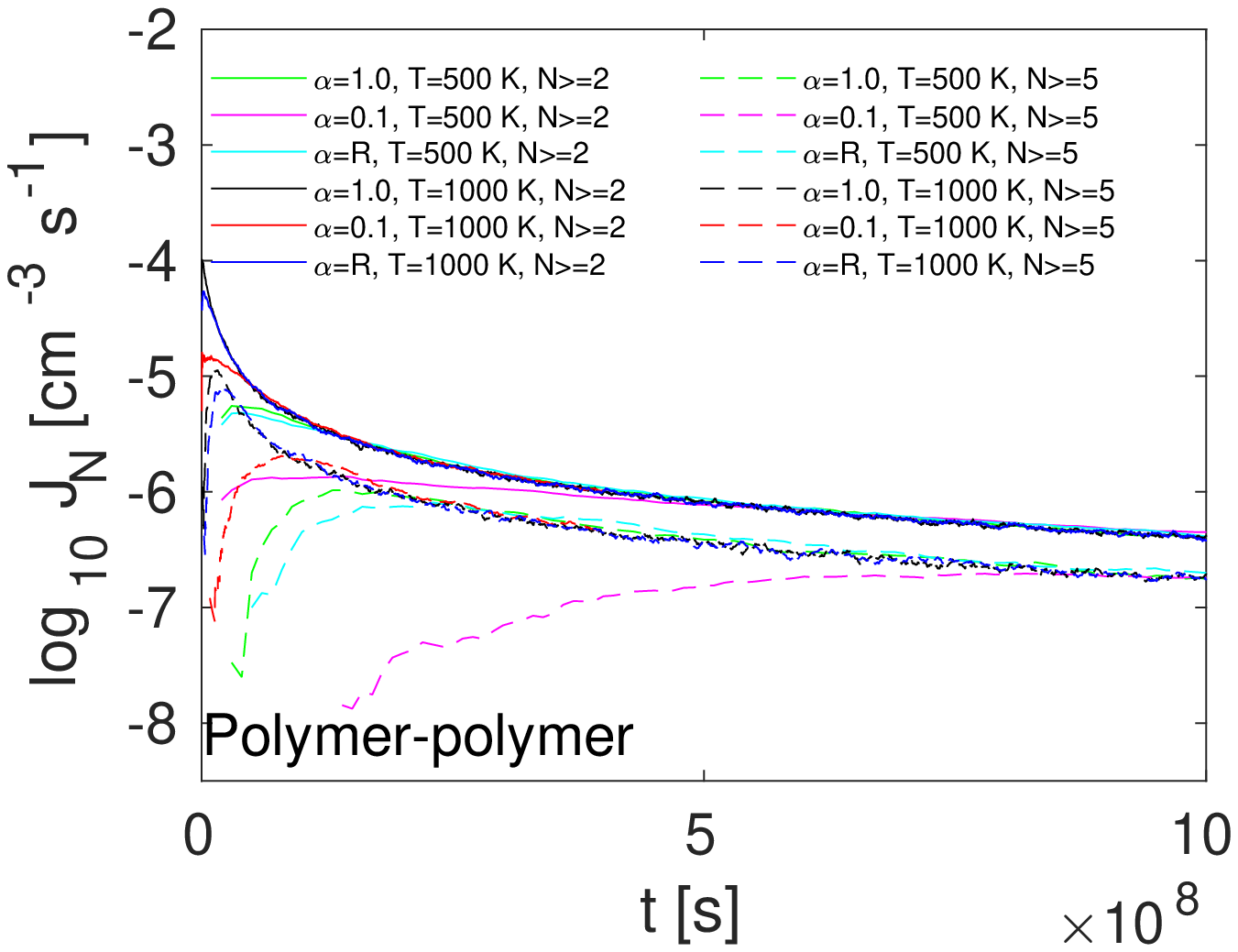}\\
\includegraphics [width=8.6cm] {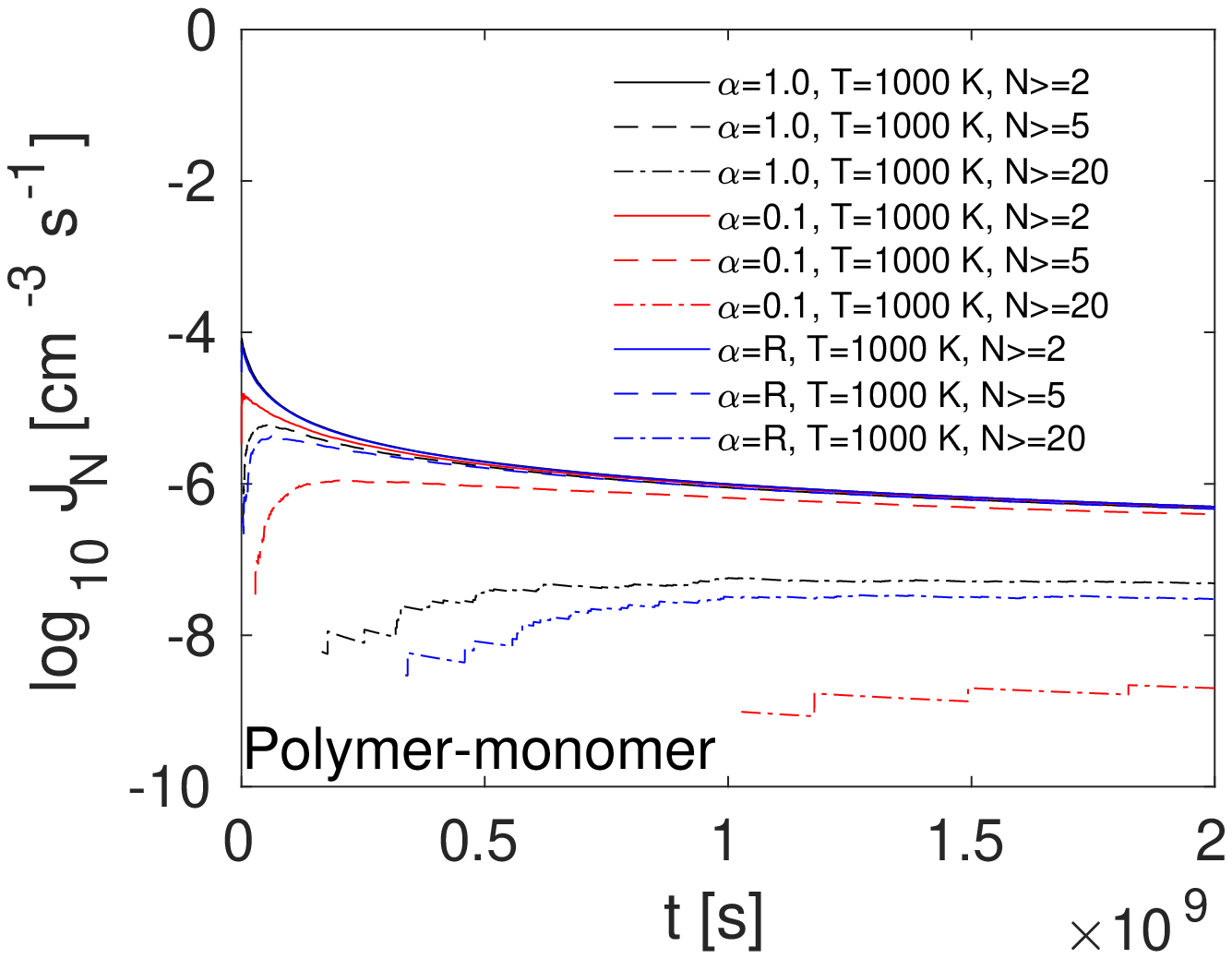} 
\includegraphics [width=8.6cm] {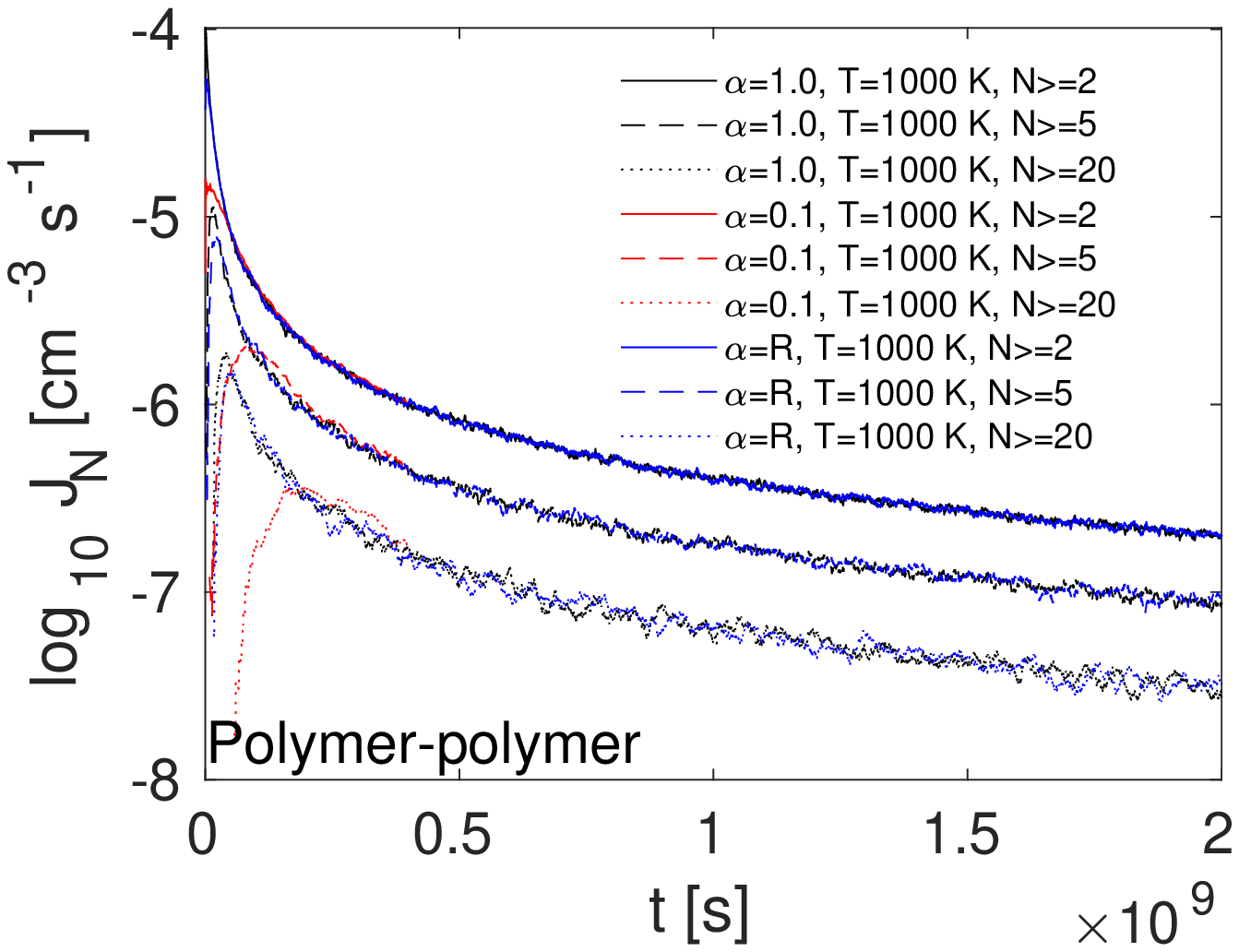} 
\caption {Time evolution of the cluster formation rate, $J_{\rm N}$ [cm$^{-3}$ s$^{-1}$], for different temperatures and sticking probabilities.\\
{\bf Left column:} For clusters larger than $N=2, 5, 20$  from polymer-monomer cluster growth. {\bf Right column:} For clusters larger than $N=2, 5, 20$  from polymer-polymer cluster growth.  {\bf Top:}  Early time evolution until $t=10^7$s for T=500K, 1000K; only $J_{\rm >2}$
and $J_{\rm >5}$ appear, {\bf Middle:} Intermediate time evolution until $t=10^9$s for T=500K, 1000K, for $J_{\rm >2}$ and $J_{\rm >5}$,   {\bf Bottom:} Long-term evolution until $t=2\cdot 10^{9}$s for $J_{\rm >2}$, $J_{\rm >5}$, and also  $J_{\rm >20}$ for T=1000K only.  No clusters $>$20 have yet formed for T=500K.}
\label {fig:nucl_rate_1}
\end {figure*}

\subsection{Cluster size distribution in a homogeneous TiO$_2$ gas: The effect of temperature and reaction efficiency}\label{s:MC_TiO2}

Figure \ref{fig:size_distr_1} shows the (TiO$_2$)$_{\rm N}$ cluster size distribution for different temperatures and different reaction efficiencies (sticking probabilities, $\alpha$) in a pure TiO$_2$ gas. The top panel shows the distributions for $N=1-50$ and $\alpha=1$ for monomer-cluster (denoted as polymer-monomer) and cluster-cluster (polymer-polymer) growth scenarios at three different times. The
middle panel shows the full size distribution for all six scenarios at 1000 K (three different values of $\alpha$ for monomer-cluster and cluster-cluster growth) at the same time ($t=4\cdot10^8 s$), while the bottom panel shows cluster size distributions at temperatures of 500, 1000, and 1500 K for both monomer-cluster and cluster-cluster scenarios for  $\alpha =1$ and $\alpha =rand$ at the same time ($t=1.7\cdot10^7 s$). These times are chosen as the highest possible time that is shared between all the shown simulations. 

In the top panel we see that the monomer-cluster scenario and the cluster-cluster scenario develop in two fundamentally different ways. The former has a Gaussian-like shape with a well-defined mean cluster diameter while the latter looks more like a power-law that reaches larger sizes than the monomer-cluster cases but also has more free monomers. When only growth via monomers is allowed, an abundance of dimers is formed as can be seen for $t=3\cdot10^7 s$. Due to the replenishment mechanism for monomers, a new monomer is created for each collision (as long as the total number of clusters is below the starting amount of 1000). So the formation of a dimer effectively consumes (at least) one monomer. This depletes the amount of monomers and makes the resulting distribution look more like the growth of an initial burst of newly formed aerosols. For the cluster-cluster cases, more large clusters are formed and these are effective at scavenging the monomers that are then replenished (nearly) 1:1. Due to the replenishing mechanism, the total cluster mass increases during the simulations at different rates for the two scenarios. In the $\alpha=1$ cases for 1000 K after $4\cdot10^8$ s seconds, there is a total amount of monomers (free + bound in clusters) of 55406 for the cluster-cluster simulation and of 8372 for the monomer-cluster simulation. We note that these are monomers and there are never more than 1000 clusters in the simulation; however, since we add monomers to the simulation domain, whenever the total particle number (monomers and clusters) declines below 1000, the total number of free and bound monomers can reach values above 1000 monomers.

The middle panel allows a comparison of the different sticking probabilities. At the shown time ($t=4\cdot10^8 s$), the distributions for all three cluster-cluster scenarios are very similar, indicating that they have reached steady-state conditions except for the very largest clusters that still continue to grow. For the monomer-cluster scenarios, the  $\alpha=0.1$ scenario is less developed than the two others, which is to be expected if a steady state has not yet been reached. Since only growth via monomers is expected, it makes sense that this scenario develops more slowly than the cluster-cluster cases. 

In the bottom panel we see the effect of temperature. For  $\alpha=1$ and monomer-polymer collisions very little has happened for 500 K whereas the 1000 K distribution has grown to larger sizes. Higher temperatures cause more collisions, which produces larger clusters. The trend of increasing concentrations going from 500K to 1000 K is similar for $\alpha=rand$ with cluster-cluster collisions allowed.
If the temperature is increased further to 1500 K, the distributions shift towards lower sizes again in both shown scenarios. This is because the evaporation increases drastically (see Fig. \ref{fig:eva}); for  $T=1000$K the evaporation probability is $10^{-9}$ to $10^{-10}$ s$^{-1}$ for small clusters while for $T=1500$ K it has increased to 1 s$^{-1}$, yielding $\approx 1000$ K as the most efficient temperature for TiO$_2$ nucleation \citep{2019MNRAS.489.4890B}. No size distributions for T=2000 K are shown since very little happens at that temperature because of the very large evaporation frequency.

Thus the formation of clusters is determined by the relation between the collision and evaporation frequencies, which are both increasing functions of temperature. If the temperature is too low nothing happens because there are too few collisions, but if it becomes too high then evaporations dominate. This is consistent with results found by \cite{2015A&A...575A..11L}.

 Figure~\ref{fig:gr_ev1} elucidates the interplay between collisions and evaporations by showing the timescale for both processes as a function of cluster size ($N$) for $T=1000$ K, $\alpha=1,$ with and without cluster-cluster growth. These are the timescales observed in the actual simulations which, for evaporation, can be compared with the evaporation frequency calculated in Fig. \ref{fig:eva}. The way the characteristic times are calculated is by looking for how long it takes for a given size $N$ to increase or decrease and averaging this over a time interval (see the figures for the exact times). Therefore, any increase in for example $N=2$ is counted as growth, but it might as well be evaporation from $N=3$. 
 
We see that the times are very close to each other for the monomer-cluster and cluster-cluster scenarios, respectively. The upper panel shows the growth and evaporation times averaged over the time until $N=20$ peaks in the cluster-cluster case (cf. Fig.~\ref{fig:nucl_rate_1} bottom right). It is obvious here that the larger sizes develop much faster for the cluster-cluster case. In the bottom panel, the cluster-cluster data is the same, but the monomer-cluster data is extended until the time where $N=20$ peaks for that scenario. We note here, like for the cluster size distributions, that the peak of the action is not at the largest nor smallest sizes but around $N=5\,\ldots\,10$. 

\subsection{Temporal evolution of formation rates}


Figure \ref{fig:nucl_rate_1} shows the formation rate (Eq. \ref{eq:MCJ*}) of clusters of different sizes in pure TiO$_2$. The top panels show the early stages of formation rates of clusters larger than size $N= 2$, $5, $ and $  20$ for times up to $1\cdot10^7s $ for T=1000 K and 1500 K and for monomer-cluster scenarios (left) and cluster-cluster (right).
The middle panel shows the same but for T= 500 K and 1000 K and extended up to $1\cdot10^9s $.
The bottom panel focuses on the T = 1000 K case and includes clusters larger than 20, with times up to $2\cdot10^9s $.
From the top panel, we notice that at 1500 K the formation rate of both cluster sizes is initially larger than for 1000 K even though we previously saw that the size distribution was further developed for 1000 K. This shows how the collision dynamics determines the cluster growth. The early part of the temporal evolution is dominated by collisions and an abundance of monomers. As time goes on fewer monomers are available, the evaporations take over, and we notice how the formation rate for $N\geq2$ becomes larger for the 1000 K case. In general there is a much larger overshoot of the formation rates for the 1500 K cases before they find a more stable level.

In the middle panels where the time frame is extended to $t=10^9$s,  we see that the 1000 K cases actually also overshoot but are slower not only to increase but also to relax. For clusters of $N\ge 2,$ the formation rates for 500K and 1000 K eventually become nearly identical while for $N\geq5$ the value for 1000 K remains significantly larger than that for 500 K in the monomer-cluster case, while in the cluster-cluster case the values find the same final level.

The lower panels allow for a detailed comparison of the different sticking probabilities at 1000 K. In the left panel (monomer-cluster cases), for $N\geq2$ there is very little difference between $\alpha=1 $and $rand$ while $\alpha=0.1 $ clearly lags behind for the early parts of the simulation. The difference between the sticking probabilities becomes clearer as we look at the larger sizes. For $N\geq20,$ the largest probability produces the largest production rate for the full duration of the simulation. 
Looking at the right hand side panel (cluster-cluster collisions), all the production rates for different sticking probabilities become similar much faster. We now understand why the size distributions in Fig.~\ref{fig:size_distr_1} are almost the same for the cluster-cluster cases and only $\alpha=0.1 $ is smaller than the others for monomer-cluster cases. While the different values of $\alpha$ affect the temporal evolution, the end result remains the same.

A general feature worth noting is that while the formation rates approach a stable value, there are fluctuations around this value. This is a feature of the MC simulations that is not captured in other models.

\begin{figure}[h!]
\includegraphics [width=9.6cm] {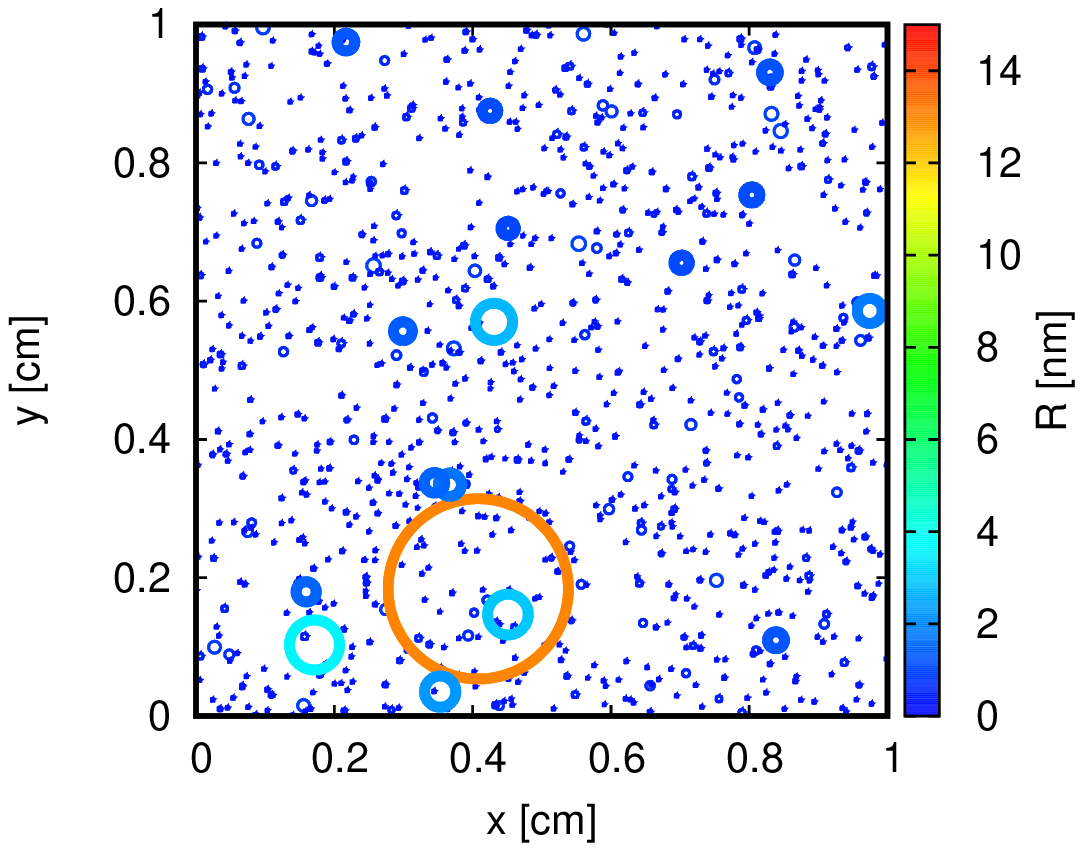}
\includegraphics [width=9.3cm] {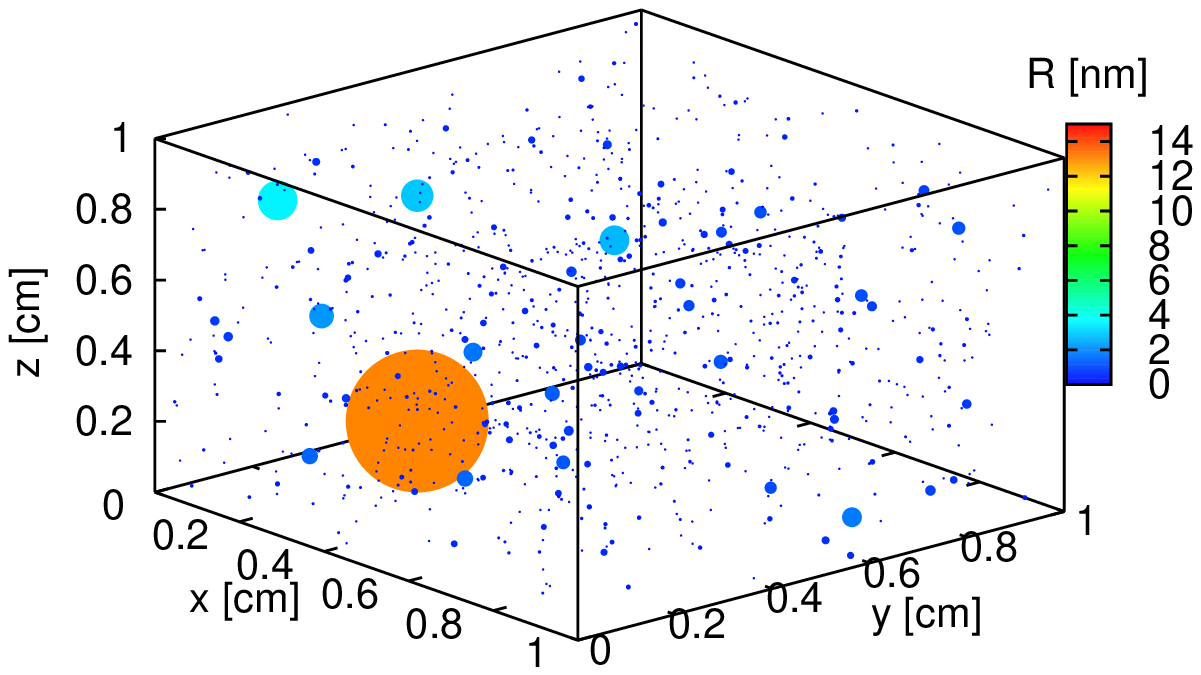}
\includegraphics [width=8.6cm] {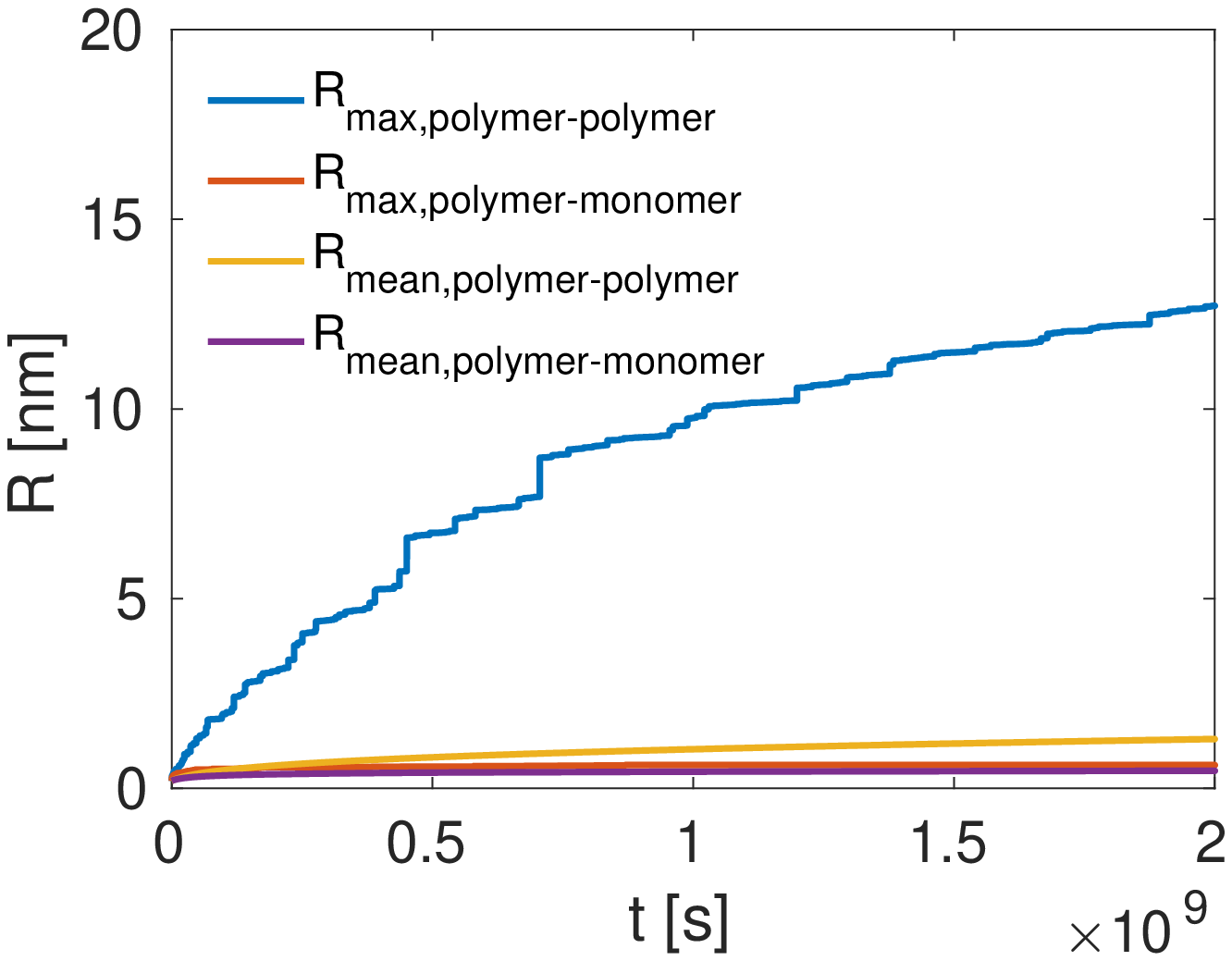}
\caption {Spatial distribution of clusters formed by polymer-polymer collisions at $t\approx 2.2\cdot 10^9$ s for $\alpha=1$ and $T=1000$ K. The largest cluster (orange) is of size $N=281755$ with R =  13 nm  (Eq.~\ref{eva.4}). Each cluster radius is amplified by a factor of $10^7$ for visualisation purposes in the plots.  {\bf Top}: Projected onto the $xy$-plane. {\bf Middle}: 3D distribution of all clusters that formed; the size is colour coded. {\bf Bottom}: Temporal evolution of maximum and mean cluster size.}
\label{fig:3d}
\end{figure}

\begin{figure*}
\includegraphics [width=9.6cm] {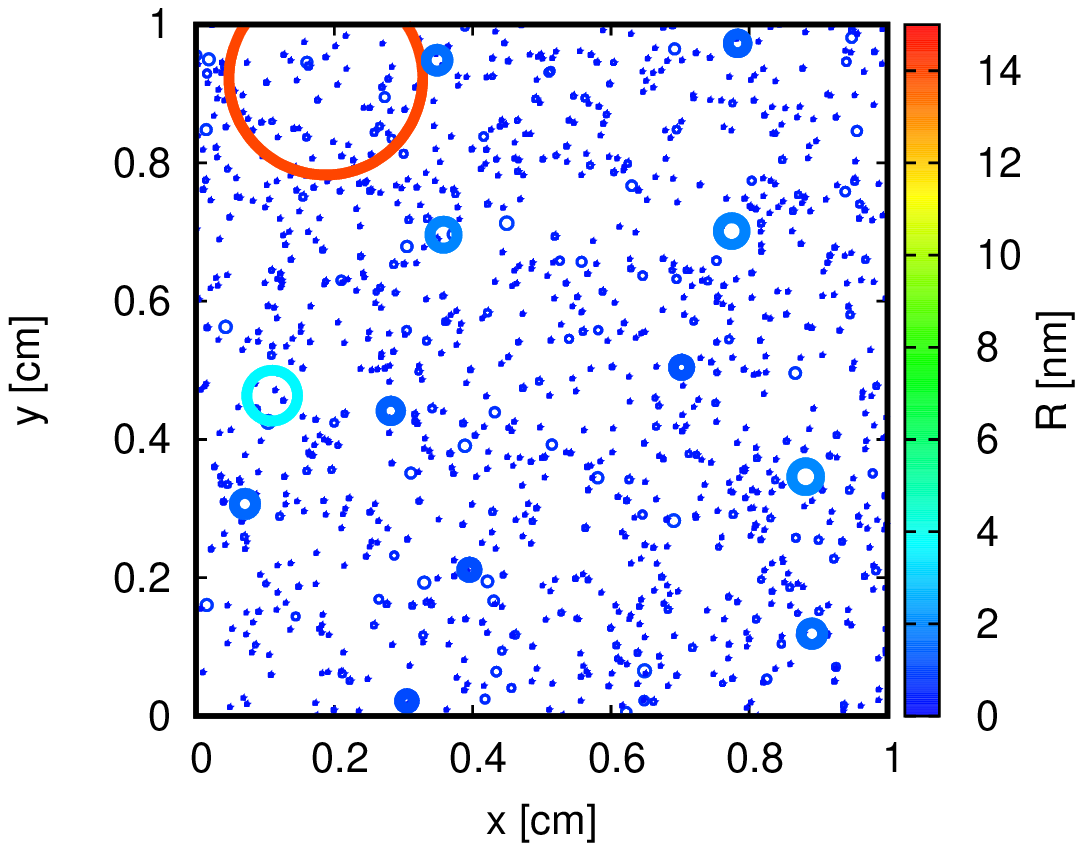}
\includegraphics [width=9.6cm] {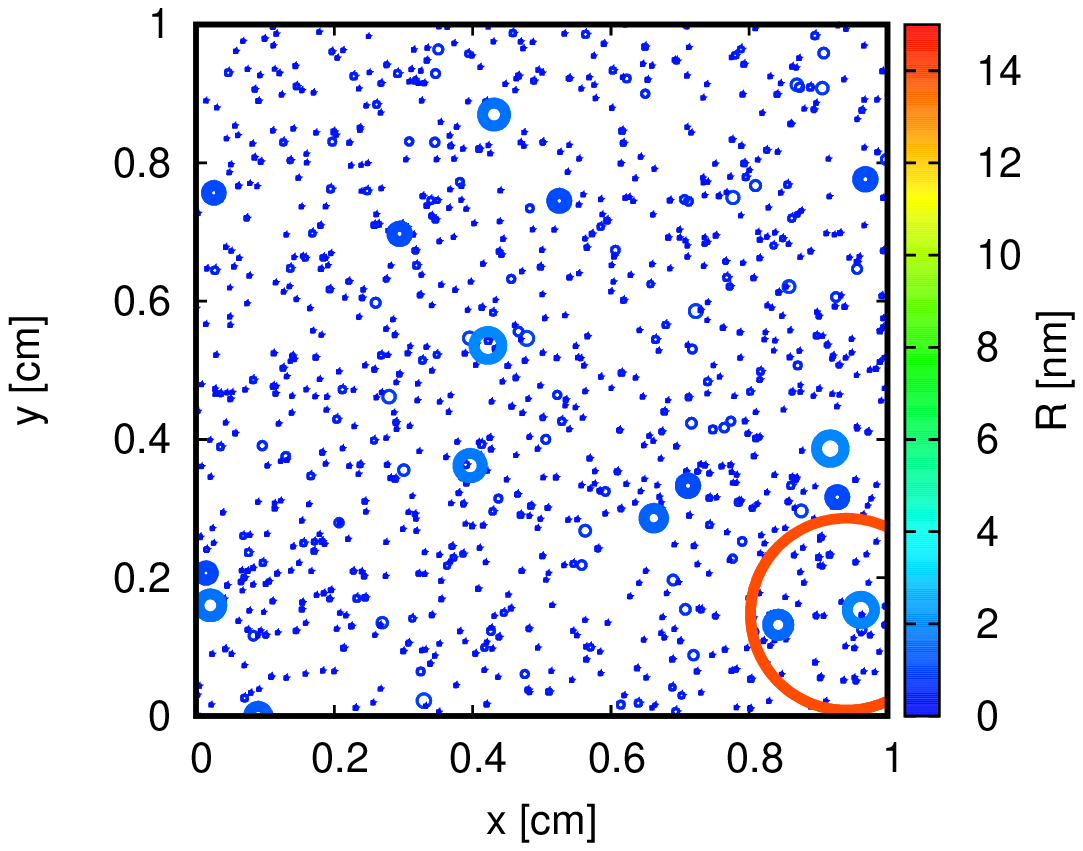}\\
\includegraphics [width=9.3cm] {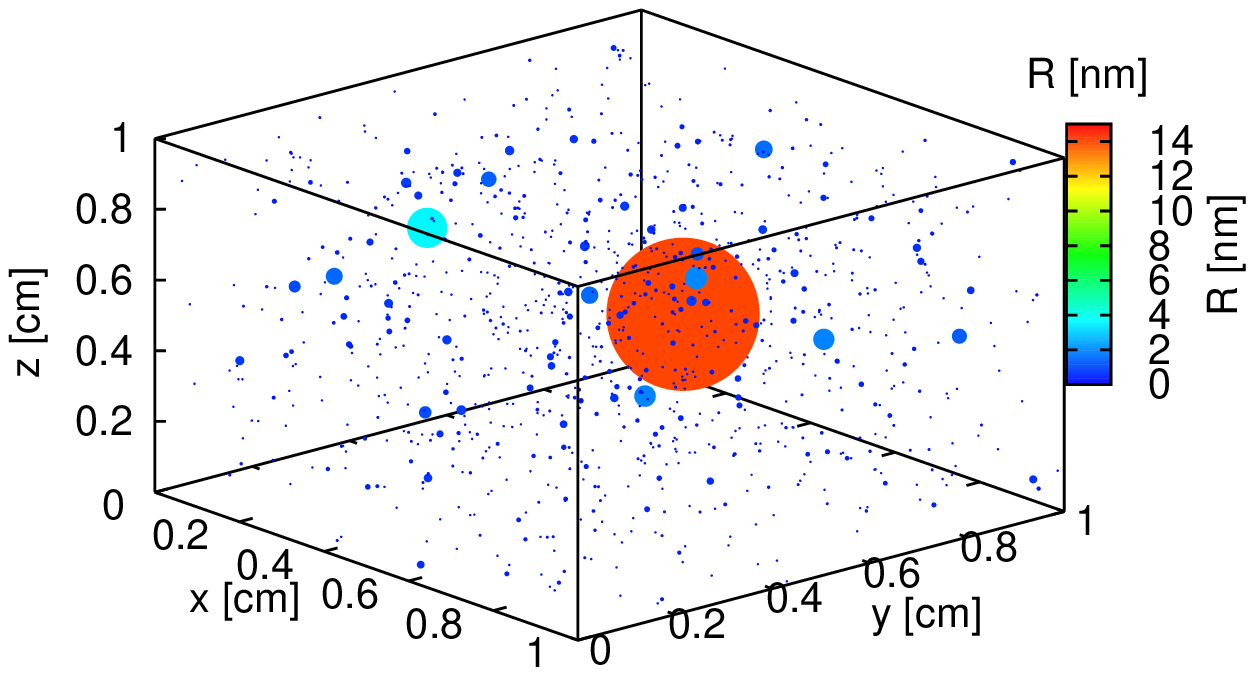}
\includegraphics [width=9.3cm] {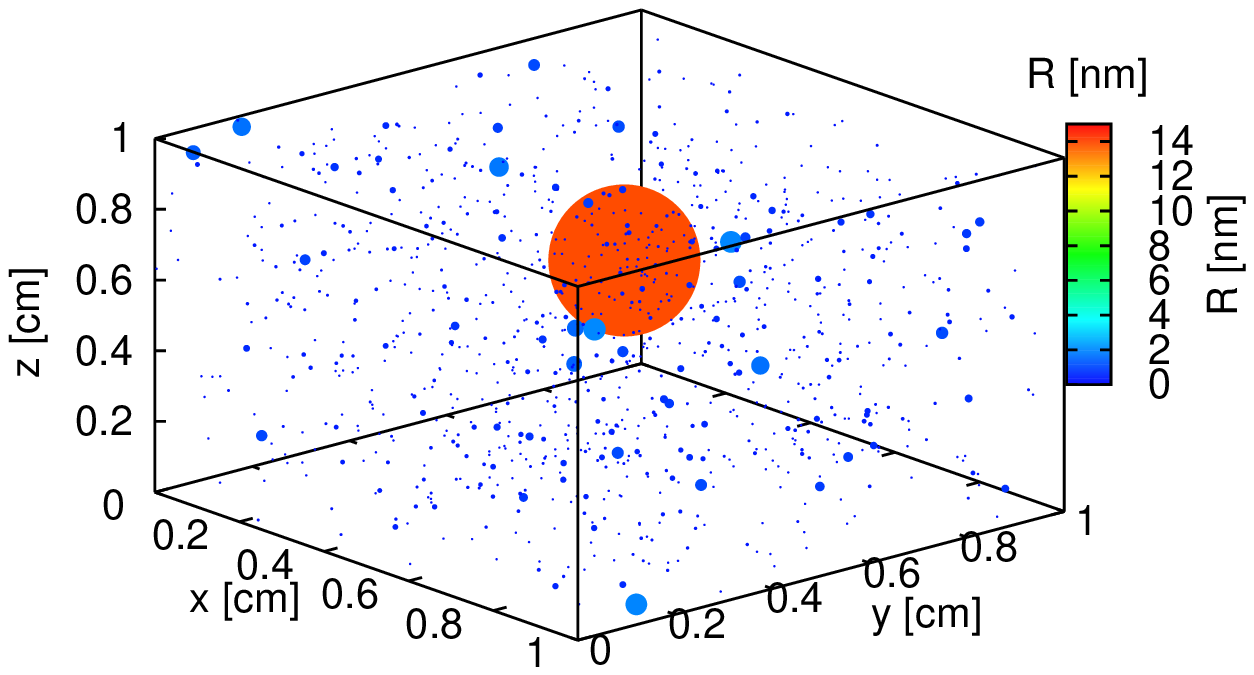}\\
\includegraphics [width=8.6cm] {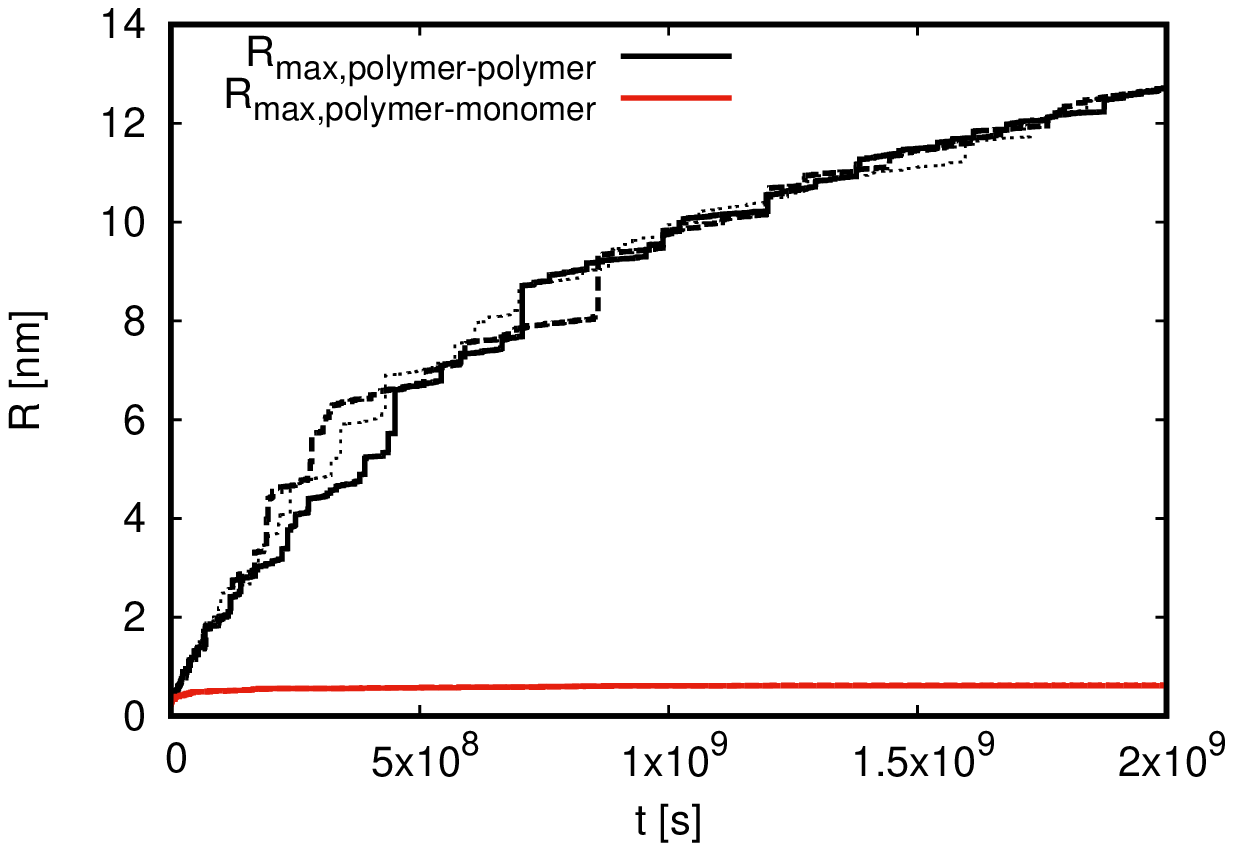}
\includegraphics [width=8.6cm] {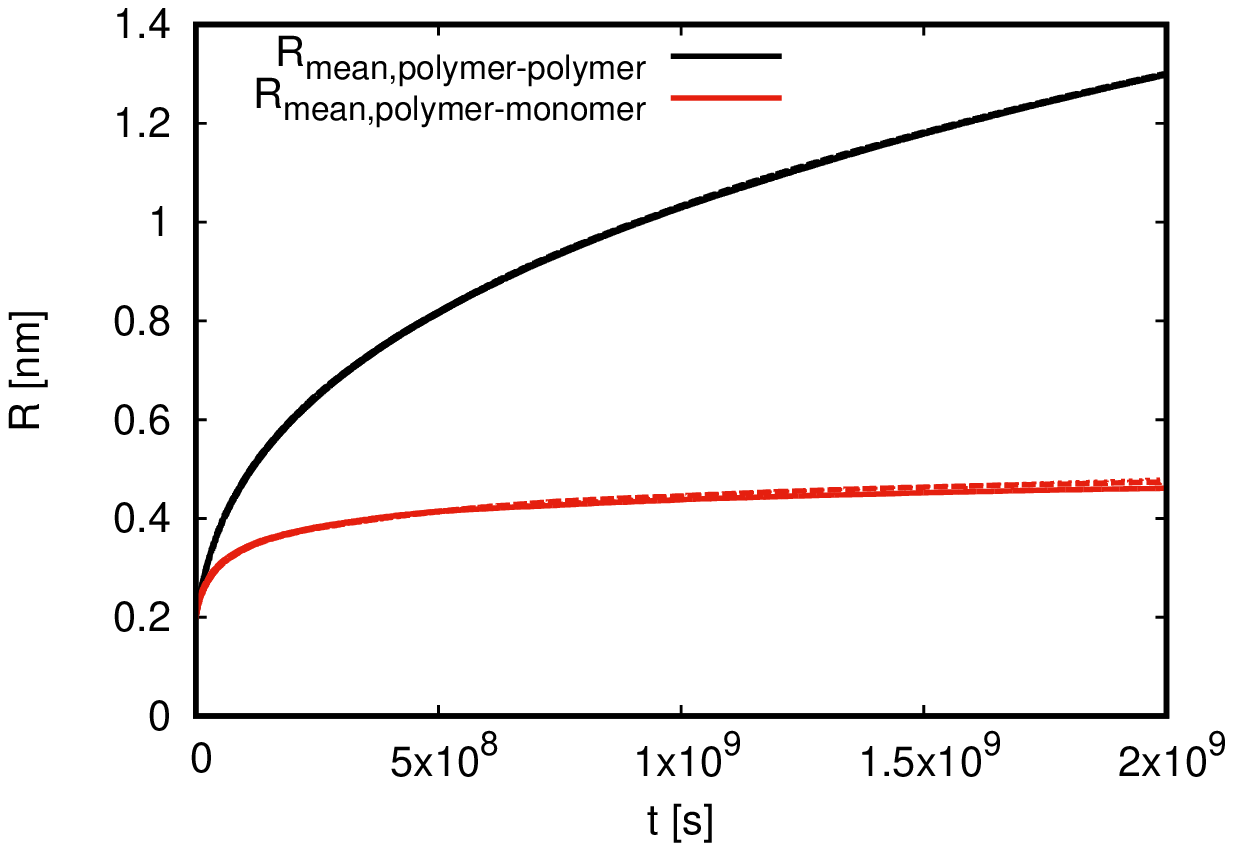}
\caption {{\bf Top, middle}: Spatial distribution of clusters at $t=2.2\cdot 10^9$ s for $\alpha=1$ and $T=1000$ K as in Fig. \ref{fig:3d}, but for two different realisations of random numbers (left and right column). {\bf Bottom}: Maximum and mean cluster size as a function of time for simulations with different seeds of random numbers (solid, dashed, dotted). The black lines in the right hand panel and the red lines in both panels almost lie on top of each other.} \label{fig:3d_2}
\end{figure*}

\subsection{Spatial distribution}\label{ss:space}

Monte Carlo simulations have shown that the cluster formation rates are affected by the stochastic nature of particle motion and particle collisions \citep{2020AerST..54.1007K}. For $\alpha=1$ and $T=1000$ K, Fig.~\ref{fig:3d} shows the spatial distribution of all clusters (top: projected onto the $xy$-plane; middle: whole 3D distribution) at $2.2\cdot 10^9$ s. It shows that most (TiO$_2$)$_{\rm N}$ clusters remain rather small but that there is one distinct cluster continuously growing (orange). Such a large cluster has a significantly large surface area, which makes it more likely to merge with other clusters in contrast to the less probable collision of two small clusters. 

The bottom panel in Fig. \ref{fig:3d} shows the mean and maximum radius of all clusters as a function of time. For the simulation allowing polymer-polymer interactions, it shows that the mean radius (orange line) hardly changes in time, limited below 2 nm until $t=2\cdot 10^9$ s, whilst the blue line depicts the creation of one big particle (orange in the upper two panels). This one distinct cluster forms very early in the nucleation process and subsequently becomes the dominantly growing cluster. The bursts in growth seen in the maximum size is due to scavenging of other large particles by the largest cluster.

The red and violet lines show the mean and maximum radius when only monomers are allowed to attach to clusters. In this case, $R_{\rm max}$ and $R_{\rm mean}$ are of comparable size and do not grow significantly until $2\cdot 10^9$ s, staying smaller than $R_{\rm mean}$ for polymer-polymer interactions.

Figure \ref{fig:3d_2} shows the same spatial distribution for $T=1000$ K and $\alpha=1$ as in Fig. \ref{fig:3d}, but for different realisations of random numbers throughout the whole simulation (e.g. checking for evaporation). It shows that the formation of one big cluster and of many small clusters is independent of the randomness of the collision process. Figure \ref{fig:3d_2} also shows the maximum cluster size (left) and the mean cluster size (right) for three different simulations with different seeds of random numbers. For all simulations, the maximum and mean cluster size show a very similar temporal evolution. The maximum radius becomes $\sim$ 13 nm after approx. $2\cdot 10^9$ s (the orange cluster in the panels above) whilst the mean radius for polymer-polymer and polymer-monomer interactions and the maximum radius for polymer-monomer interactions is limited to $\lesssim 2$ nm.

\begin{figure}
  \hspace*{-0.2cm}   \includegraphics[width=9.5cm]{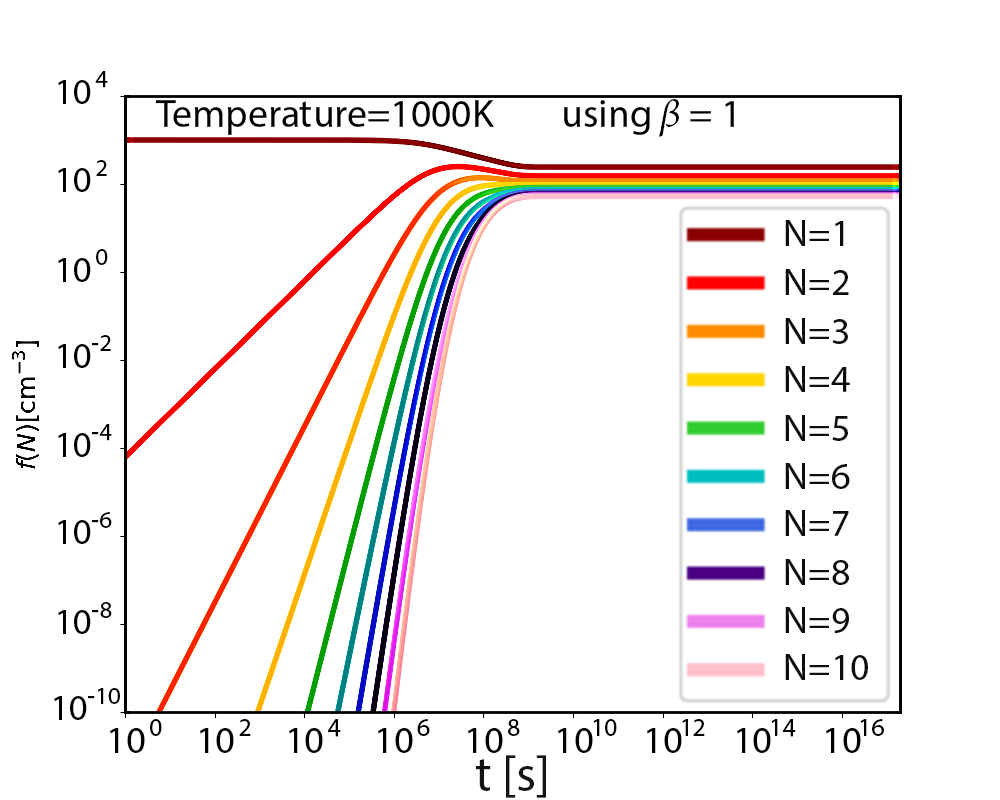}
\includegraphics[width=8.7cm]{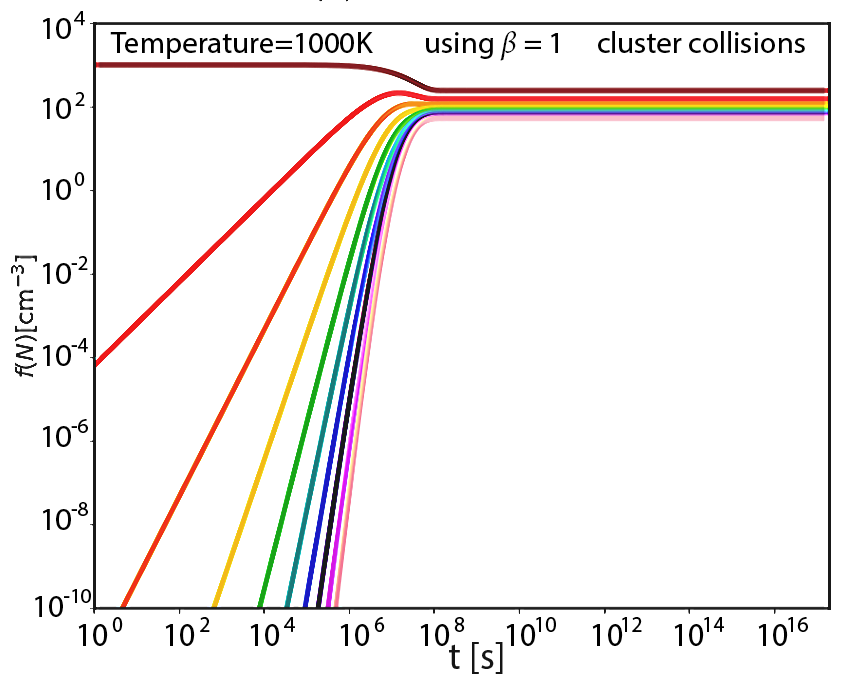}
 \includegraphics[width=8.7cm]{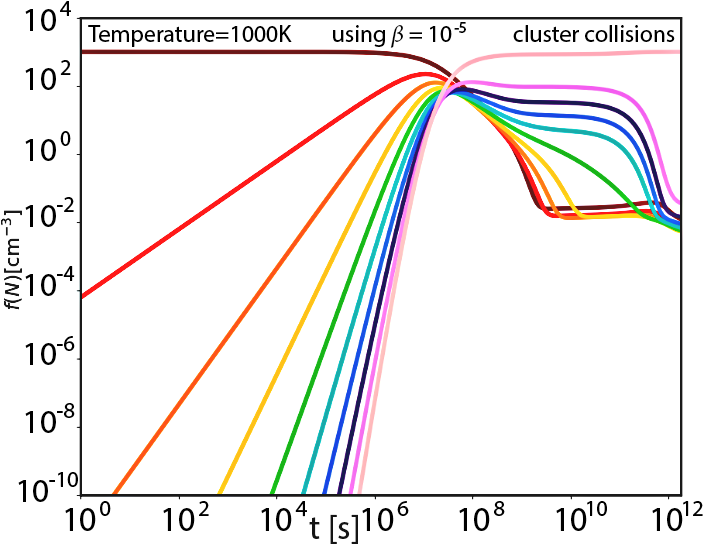}
\caption{Cluster size distributions $f(N)$ [cm$^{-3}$] as a result of  kinetic approach based on thermodynamic cluster data for clusters sizes $N=1\,\ldots\,10$ for  $\alpha=1$ for $T=1000$K,  $\beta=1.0$ for the top two panels, and $\beta=10^{-5}$ for the bottom panel in order to visualise the relative importance of cluster evaporation. The differences between the cluster-monomer (top) and the cluster-cluster (bottom) results appear small for the small clusters, and are amplified if the evaporation efficiency is suppressed.} 
    \label{fig:fn_kin1}
\end{figure}

\section{Comparing Monte Carlo and kinetic results for TiO$_2$ cluster abundances}\label{s:comp}

Figure~\ref{fig:fn_kin1} shows the cluster size distributions based on our kinetic rate equation approach (Eq.~\ref{eq:lma}, Sect.~\ref{ss:kinap}) for the cluster sizes $N=1\,\ldots\,10$. These results utilise the thermodynamic cluster data as presented in \cite{2015A&A...575A..11L}, which are based on the set of (TiO$_2$)$_{\rm N}$ cluster structures presented by \cite{2000JPhB...33.3417J}.
Combining Eqs.~\ref{eq:master} and~\ref{eq:clflux}  results in 
\begin{eqnarray}
\label{eq:cc}
    \frac{df(N,t)}{dt} = \sum^{N-1}_{i=1}\frac{f(N-i)}{\tau_{\rm gr}(N-i)} - \frac{f(N)}{\tau_{\rm ev}(N)}\\
       \label{eq:cc2}
        \frac{df(N+1,t)}{dt} = \sum^{N-1}_{i=1}\frac{f(N)}{\tau_{\rm gr}(N)} - \frac{f(N+i)}{\tau_{\rm ev}(N+i)}.
\end{eqnarray}
This system of ordinary differential equations (ODEs) allows us to consider the effect of cluster-monomer growth (for $i=1$) and cluster-cluster  nucleation, similar to  Sect.~\ref{s:MCTiO2}. All variables have been defined in Sect.~\ref{ss:kinap}. 

Equations~\ref{eq:cc} and~\ref{eq:cc2} are solved for a temperature of $T=$1000K. The initial monomer density of $f_{\rm TiO2}=f(1)=10^3$cm$^{-3}$ is chosen in accordance with Fig. 4 in \cite{2015A&A...575A..11L}, which is equal to the initial particle density $f_{\rm tot}$ used in our MC simulations. It is a representative value for the gaseous TiO$_2$ abundance in the atmospheric regions where $T=1000\,\ldots1500$K (and the gas pressure $p\approx 10^{-6}\ldots 10^{-4}$ dyn/cm$^2$ for a model atmosphere of $T_{\rm eff}=$1600K, log(g [cm s$^{-2}$])=3)  where TiO$_2$-seed formation can be typically expected in brown dwarfs and gas giant exoplanets. Higher TiO$_2$ number densities may occur for higher gas densities. As we focus here on the comparison of two models, the kinetic nucleation and the MC approach, and since MC simulations usually take a significant amount of time \citep{2018JCoPh.363...30K}, here we offer one example. In the future, we plan to extend our comparison to other model atmospheres for example $f_{\rm tot}=10$ or $f_{\rm tot}=10^8$ cm$^{-3}$.

\subsection{Results from thermodynamic cluster formation modelling}

Analogously to the 3D MC simulations in Sect.~\ref{s:MCTiO2}, we explore cluster growth by cluster-monomer and by cluster-cluster process for the first ten (TiO$_2$)$_{\rm N}$ clusters. For $T=1000$K, the solution of Eqs.~\ref{eq:cc} and~\ref{eq:cc2} is shown in Fig.~\ref{fig:fn_kin1}. As previous work \citep{2019MNRAS.489.4890B} and our MC  approach have demonstrated for the considered cases, the growth of TiO$_2$ clusters is more dominant than the evaporation frequency at 1000 K, which is why we focus on this temperature in the following. Both cases look rather similar, except for the larger cluster sizes that appear earlier if cluster growth proceeds via cluster-cluster collisions. 
In order to understand the importance of cluster evaporation, we have solved the kinetic model for negligible evaporation ($\beta=10^{-5}$) such that the growth of clusters is the only relevant process.
Large and size-dependent differences occur regarding the relaxation timescale to a steady state if the evaporation probability is decreased (i.e. suppressed) uniformly for all clusters (Fig.~\ref{fig:fn_kin1}, bottom panel). The artificial suppression of cluster evaporation demonstrates that larger clusters become increasingly more stable against destruction compared to the smaller clusters, driving the cluster size distribution away from an equilibrium distribution. We conclude that the maximum cluster size of $N=10$ monomers is still too small for the divergence of the monomer-cluster and the cluster-cluster growth to appear as pronounced as in our 3D MC simulations.

\begin {figure}
\includegraphics [width=9.3cm] {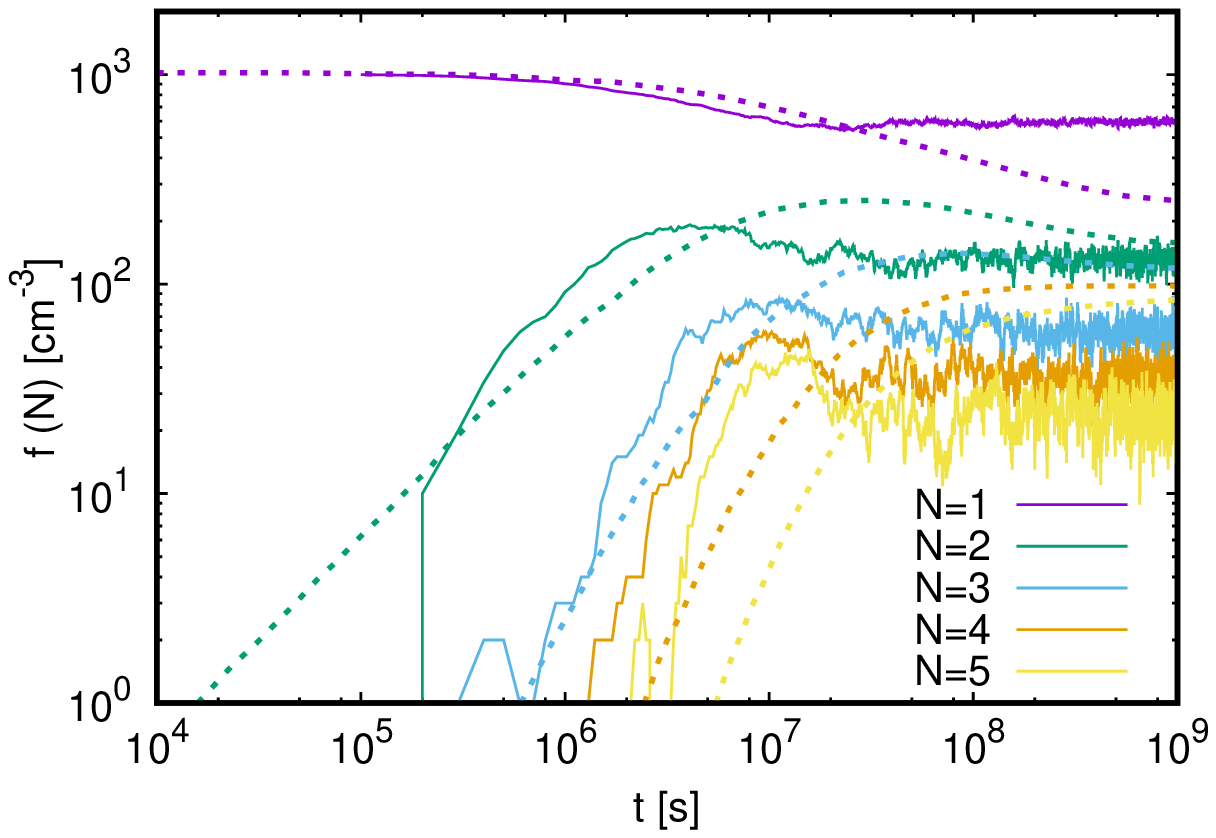}
\includegraphics [width=9.3cm] {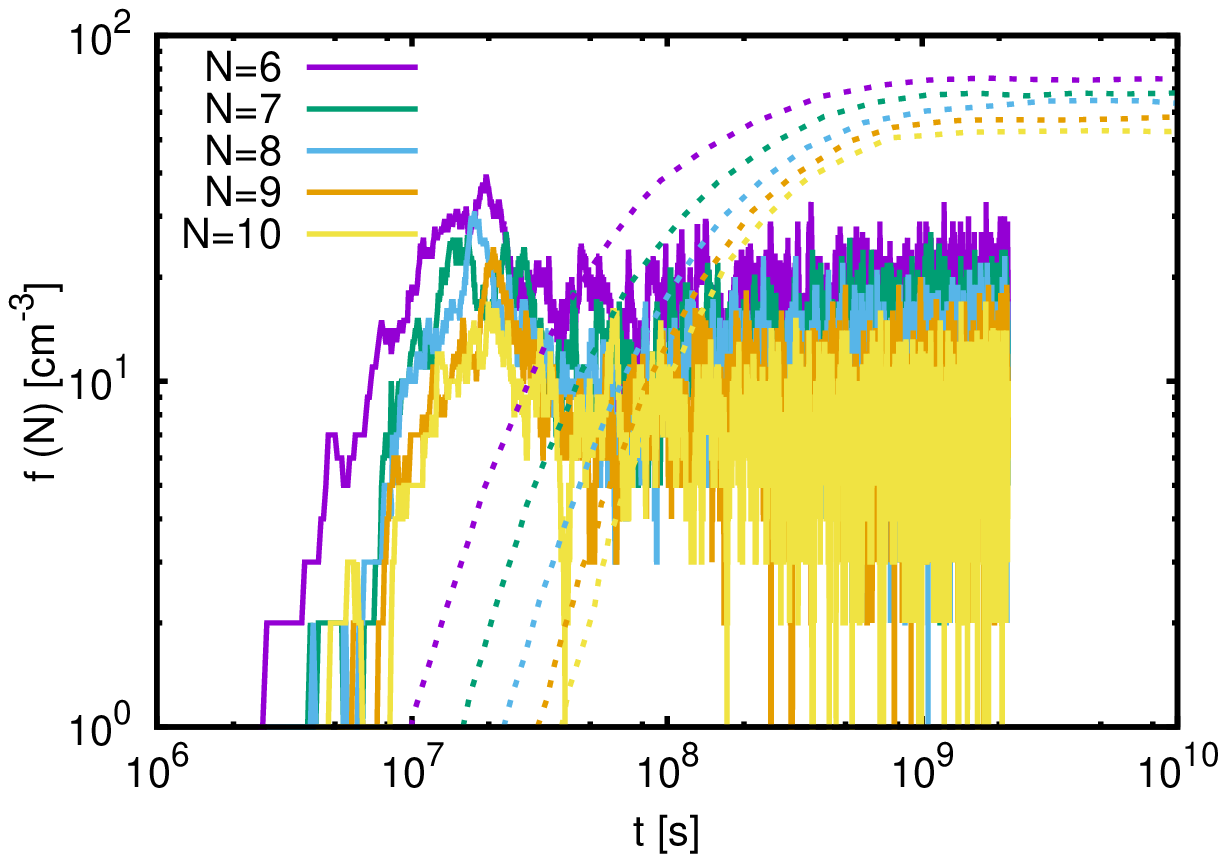}
\caption {Comparison of the cluster number densities for  $N=1\ldots\,5$ (top) and  $N=6\ldots\,10$ (bottom) resulting from the 3D Monte Carlo (3D MC, solid) and from the kinetic approach (dotted) for $\alpha=1$, $\beta=1$ and $T=1000$K. Deviations are largest for the larger clusters as the kinetic approach is limited to N=10 as the maximum but the MC approach can grow to N>10, hence depleting the smaller cluster numbers.}
\label{fig:MCTDcomp}
\end {figure}

\subsection{Comparison of 3D MC and thermodynamic cluster formation modelling}

Our study of (TiO$_2$)$_{\rm N}$ cluster formation as a precursor for nucleation seed formation in exoplanet and brown dwarf atmospheres, as well as in AGB star envelopes, points towards considerable timescales before a steady-state solution can be reached. The timescale shortens if all possible clusters participate into the growth process, instead of cluster growth by monomer addition only. This finding is consistently confirmed by both approaches, the  molecular-dynamics 3D MC and thermodynamic cluster formation modelling as shown in Figs.~\ref{fig:nucl_rate_1} and \ref{fig:fn_kin1}.
\cite{2020AerST..54.1007K} have further demonstrated that a decreasing gas density does not considerably affect the timescale on which the steady-state formation rate is reached, and that the gas density does not affect  the value of the steady-state formation rate. The study by \cite{2020AerST..54.1007K} addresses H$_2$SO$_4$-H$_2$O cluster growth for Earth atmosphere conditions, however. 

A comparison between our two methods can only be carried out for the cluster sizes $N=1\,\ldots\,10$ because of the presently available thermodynamic data for the kinetic approach. For $T=1000$K, we compare the cluster-cluster results only. Figure~\ref{fig:MCTDcomp} separates the comparison for the small cluster ($N=1\,\ldots\,5$) and the intermediate cluster sizes ($N=6\,\ldots\,10$) because of their qualitative differences. Generally, the cluster number densities for $N=1\,\ldots\,10$ and their time evolution agree reasonably well between the 3D MC  and the kinetic nucleation approach. Both reach the steady-state values after a considerable time, hence there is an individually constant number density for all clusters shown. Both approaches predict the small clusters to be the most abundant and the largest cluster sizes to be the least abundant. Figure \ref{fig:MCTDcomp} shows that the density $f(N=10)$ of 10-mers after one year $\sim 3\cdot 10^7$ s is on the order $10^1$ cm$^{-3}$; using an improved nucleation theory, which abandons equilibrium assumptions, discards growth restrictions, and uses quantum mechanical properties, \citet{2019MNRAS.489.4890B} find $f(10)\approx 10^2-10^3$ cm$^{-3}$ after one year for AGB wind conditions for an initial mass density $\varrho_{\textnormal{Ti}}\approx 2.84\cdot 10^{-15}$ kg m$^{-3}$. As their $\varrho_{\textnormal{Ti}}$ is approximately 35.5 times larger than $\varrho_{\textnormal{Ti}}\approx 8\cdot 10^{-17}$ kg m$^{-3}$, as used in this work, we see that their values agree well with the values obtained by our simulations. We also note that the nucleation rate of clusters does not depend linearly but rather superlinearly on the initial particle density \citep{Dunne2016,2018JCoPh.363...30K}.

Although the agreement between the MC and the kinetic approach for $T=1000$K may seem a simple observation to make, it is a very important result to emphasise as both methods approach the problem from rather different angles. The characteristic material properties are only described by an evaporation frequency in the 3D MC  (Eqs.~\ref{eq:eva.1} to  \ref{eva.4}) whereas the kinetic nucleation approach requires individual thermodynamic properties for each cluster size (and ideally also their isomers; Eq.~\ref{eq:dG}). However, the simplicity of the 3D MC approach is offset by the immense computational time demand for a system as small as $10^3$ particles in 1 cm$^3$.

For the case of $\alpha=1$ and $\beta=1$, the individual cluster number densities do not vary by more than a factor of 5 between 3D MC and the kinetic approach in the steady state. As Fig. \ref{fig:size_distr_1} shows, there is no significant difference in the evolution of the size distributions and thus of the cluster number densities when using $\alpha=rand$ instead of $\alpha=1$; for $\alpha=0.1$, the temporal evolution of the number densities is delayed by a factor of $\sim 10$ compared to $\alpha=1$. All 3D MC cluster number densities are lower in steady state than the kinetic results, except for $N=1$ and $N=2$. The  3D MC monomer density ($f(N=1)$) fluctuates around a nominal value that results from the condition of keeping the total number of particles within the 3D MC computational domain constant. The divergence of the 3D MC monomer density (solid purple line, top panel in Fig.~\ref{fig:MCTDcomp}) from the kinetic nucleation values (dashed purple line) occurs when other cluster sizes become increasingly more abundant. That is when the 3D MC code replenishes the monomers such that the total number density is kept constant to the value of $10^3$ cm$^{-3}$.

All 3D MC cluster number densities in Fig.~\ref{fig:MCTDcomp}  are lower in steady steady than the kinetic nucleation results (expect for $N=1, 2$) because the 3D MC simulations offer reaction paths to larger cluster sizes $N>10$ whereas the kinetic model as formulated in Eqs.~\ref{eq:cc} and \ref{eq:cc2} is limited by $N=10$. Therefore, the growth of larger clusters  of $N>10$ will deplete the number of smaller clusters of $N<10$ in the 3D MC results, but not in the kinetic nucleation results. All 3D MC cluster number densities in Fig.~\ref{fig:MCTDcomp} for $N>1$ overshoot the kinetic nucleation values during the beginning of the nucleation process at $t<10^{7.5}$s, meaning they form more efficiently at earlier times than with the kinetic approach.
That clusters appear faster with the 3D MC approach than with the kinetic nucleation could be because the randomness of the MC code produces fluctuating collision rates that deviate from the average values used by the kinetic nucleation code. Since evaporation is so dominant for the smallest sizes, faster growth due to an above average number of collisions can result in faster growth rates for small clusters, as shown by \cite{olenius2018} and \cite{ 2020AerST..54.1007K}.

Another reason could be 3D effects in the MC simulations due to local density inhomogeneities (as seen in Fig. \ref{fig:3d}), which are not seen in a kinetic modelling approach.
However, this finding does suggest that processes that introduce an intermittent density distribution of the gas-phase may have the same effect on the cluster formation rate early on in the nucleation process. Such a process is turbulence and its amplifying effect for chemical reactions has been demonstrated in engineering as well as in cloud formation modelling. \cite{2004A&A...423..657H} simulated turbulent cloud condensation by modelling driven turbulence on mesoscopic scales in brown dwarf and exoplanet atmospheres. It was demonstrated that the efficiency of seed formation increases in a turbulent compared to a laminar gas, and that the formation of CCNs  may occur in an event-like fashion. The amplifying effect of a turbulent fluid field on chemical processes can generally be understood as a turbulent fluid field creating an increased reactive surface compared to a laminar flow, an effect that has been demonstrated for thermonuclear deflagration in supernovae (\citealt{2005CTM.....9..693S}). Another effect that can produce this kind of overshoot is a large injection of monomers, such as when a host star rises on a planet. An initial burst of small clusters is then dampened as larger clusters start to scavenge them or the monomer concentration is depleted.

\begin{figure}
\includegraphics [width=9.3cm]{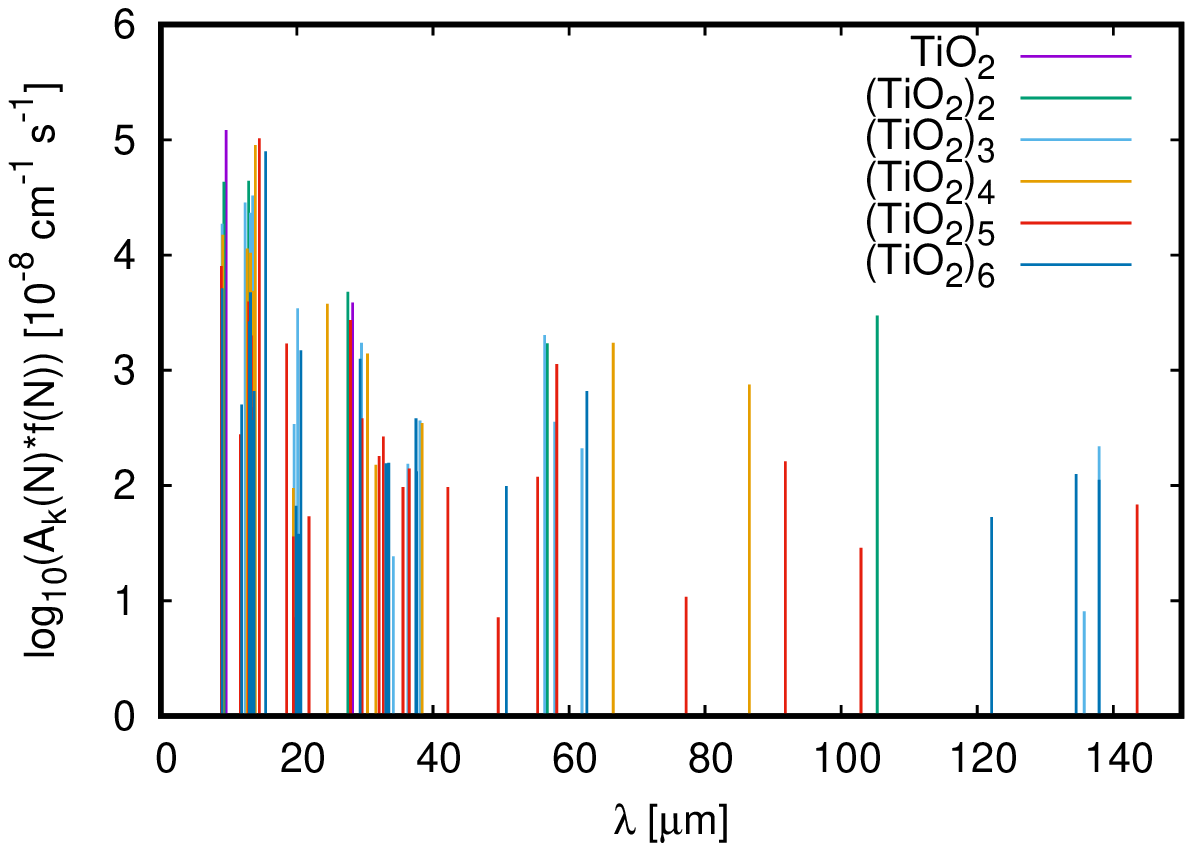}
\includegraphics [width=9.3cm]{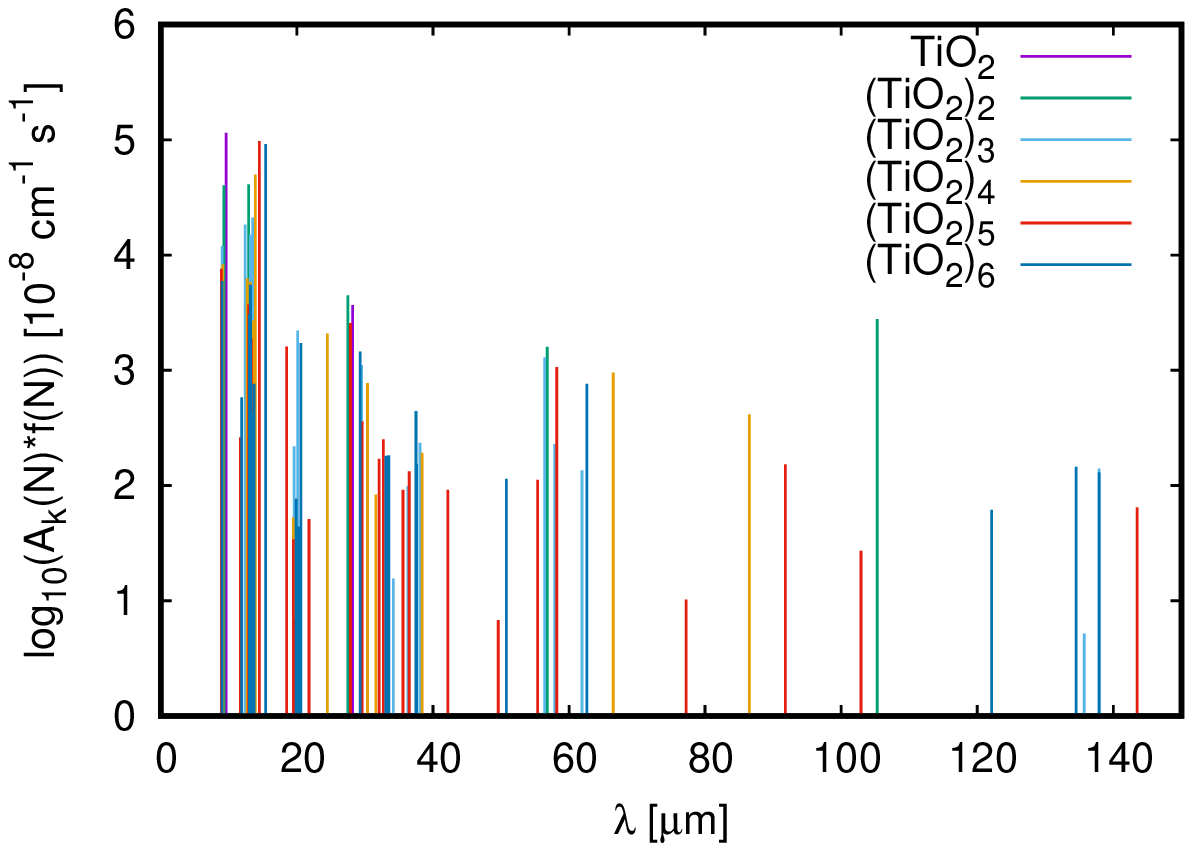}
\caption {Line opacity, $A_{\rm k}(N)\cdot f(N)$ [$10^{-8}$ cm$^{-1}$ s$^{-1}$], for (TiO$_2$)$_{\rm N}$ clusters $N=1\,\ldots\,6$ in the wavelength window relevant for potential JWST observations. The (TiO$_2$)$_{\rm N}$ integrated line absorption coefficient, $A_{\rm k}$ [$10^{-8}$ cm$^2$s$^{-1}$], data is  from Appendix C in \cite{JeongPhd}. The 3D MC cluster number density, $f(N)$ [cm$^{-3}$], is used for $t=10^7$s (top panel) and $t=10^9$s (bottom panel). Both times show little difference in  opacity despite their different cluster number densities (Fig.~\ref{fig:MCTDcomp}). (TiO$_2$)$_{\rm 4}$, (TiO$_2$)$_{\rm 6}$  , and (TiO$_2$)$_{\rm 6}$ have the largest line opacity amongst the $N=1\,\ldots\,6$ clusters. Their detailed line spectra are shown in Fig~\ref{fig:Ak_Jeong}.}
\label{fig:Ak_Jeong0}
\end {figure}

\begin{figure*}
\includegraphics [width=9.3cm]{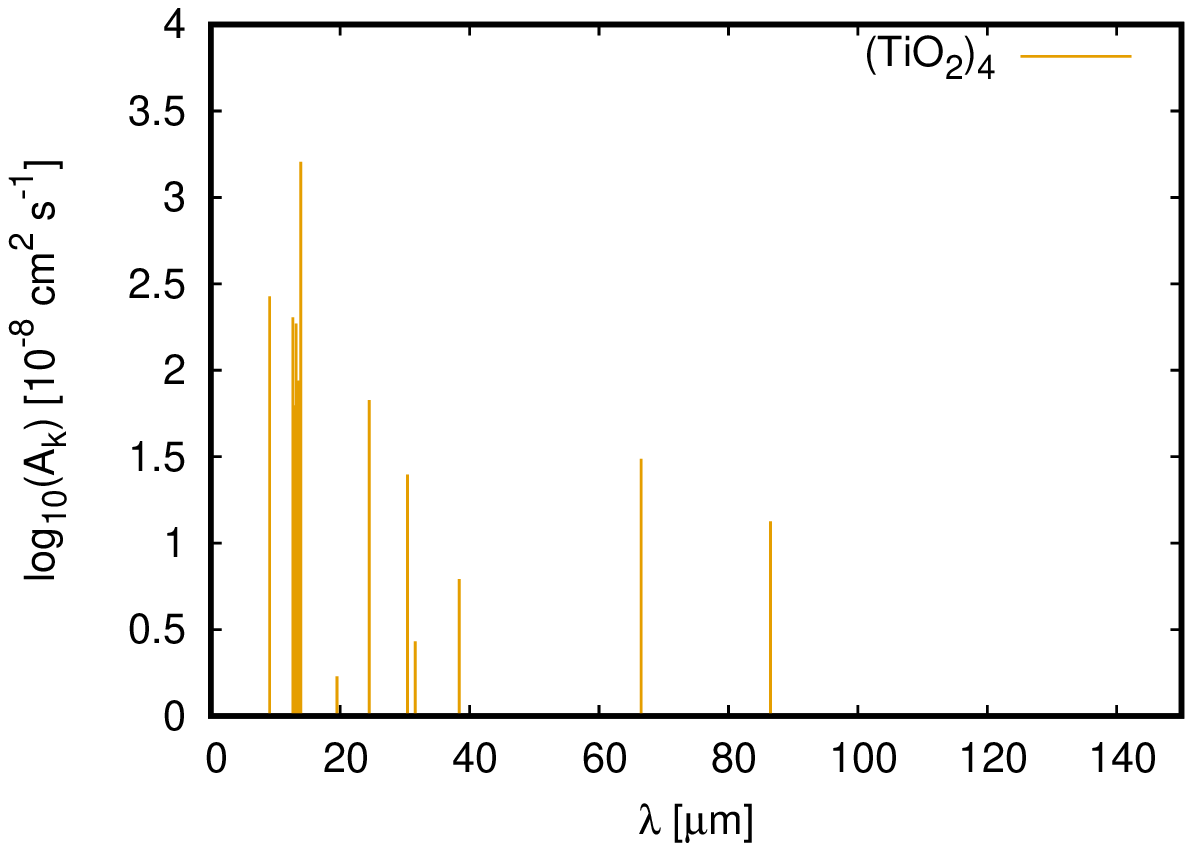}
\includegraphics [width=9.3cm]{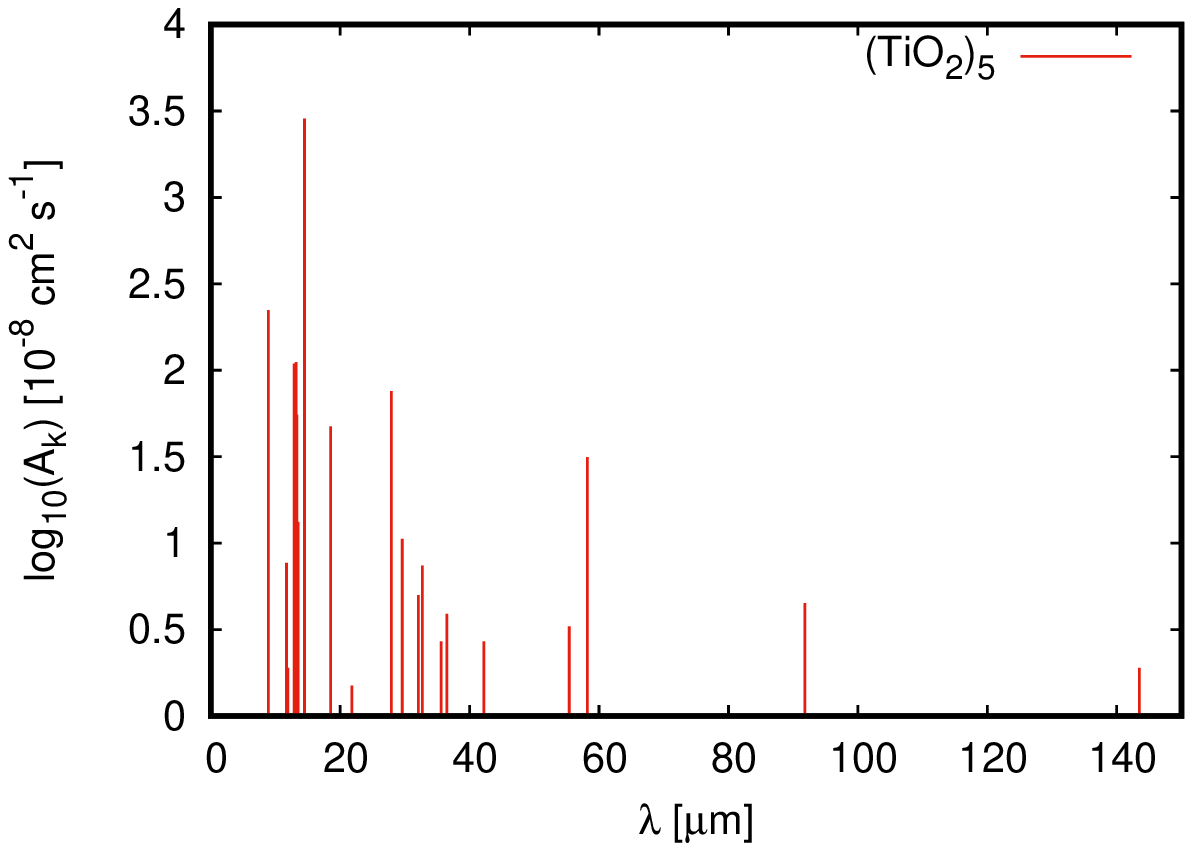}\\
\includegraphics [width=9.3cm]{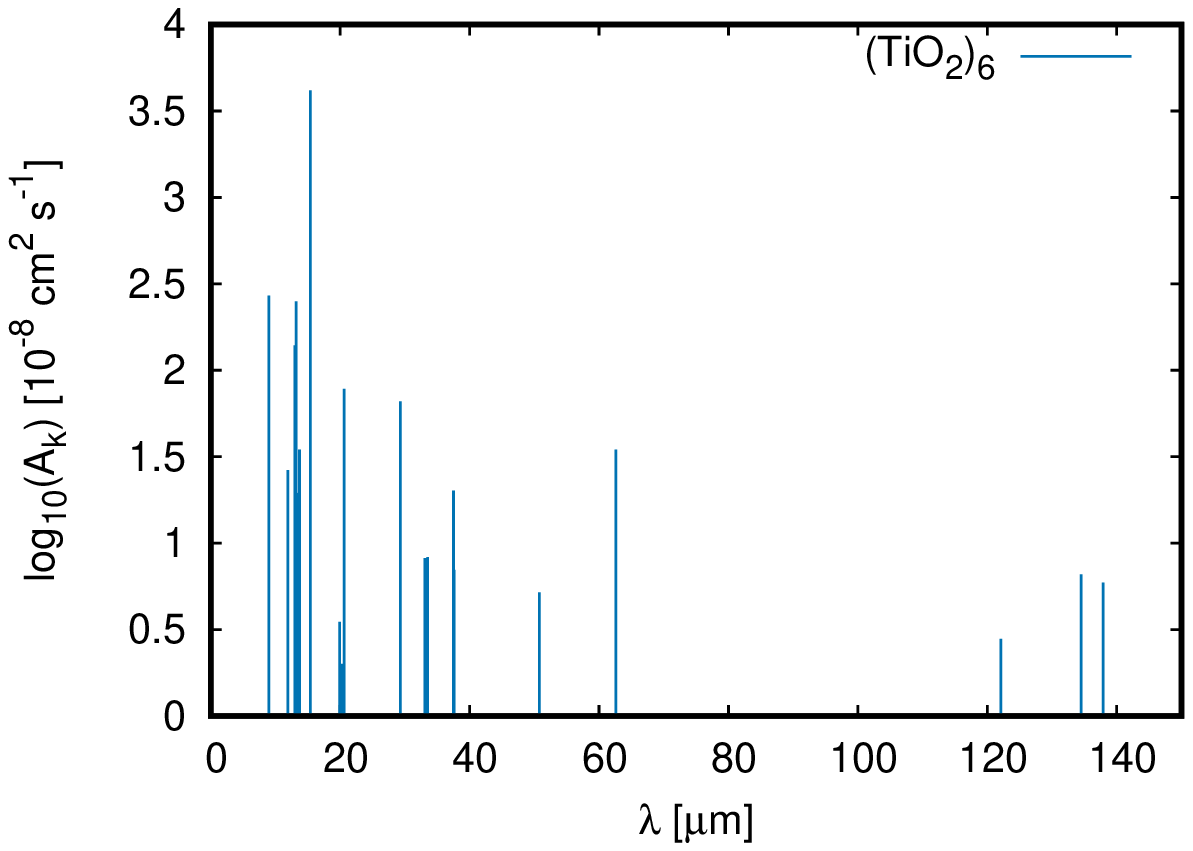}
\includegraphics [width=9.3cm]{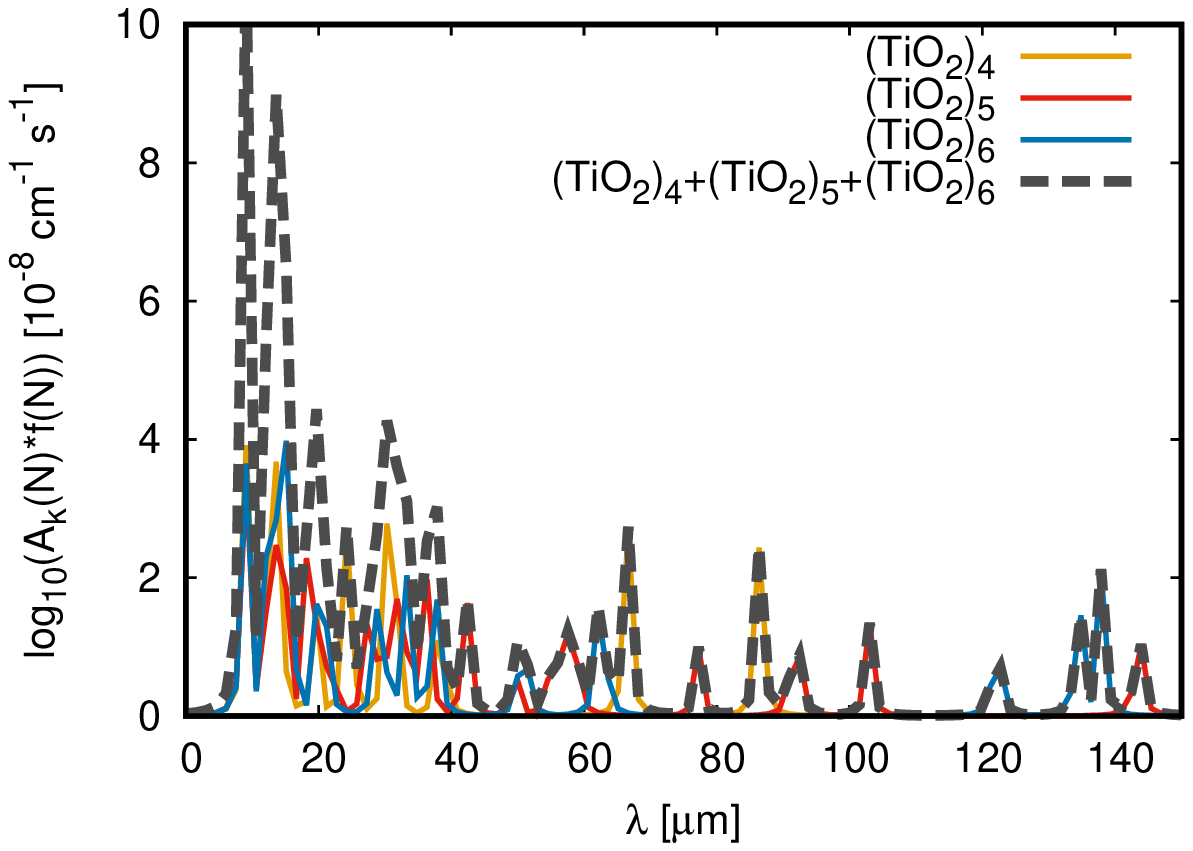}
\caption {Cluster line opacities. (TiO$_2$)$_{\rm N}$ clusters for $N=4, 5, 6$ have the strongest IR vibrations line absorption coefficient, $A_{\rm k}$ [$10^{-8}$ cm$^2$s$^{-1}$], according to  \cite{2000JPhB...33.3417J} (their Eq. 3)  amongst their TiO$_2$ cluster ensemble of $N=1\ldots6$. The data is reproduced from Appendix C in \cite{JeongPhd}. The Lorentz profile folded lines of (TiO$_2$)$_{\rm N}$ $N=4, 5, 6$  are shown in the lower right panel for $t=10^9$s.}
\label{fig:Ak_Jeong}
\end {figure*}

\section{Discussion} \label{s:discussion}

We have explored the formation of (TiO$_2$)$_{\rm N}$ clusters as precursors of cloud formation using a 3D MC approach and a thermodynamic rate equation approach in order to validate both approaches and to test assumptions that are needed on both sides. The resulting cluster number densities are promisingly comparable within a factor of 5. This conclusion is limited to the cluster sizes for which thermodynamic data are available ($N=1\ldots 10$). 

Both approaches demonstrate that the time to approach a steady-state cluster size distribution is substantial for a monomer number density that is reasonable to expect for exoplanet and brown dwarf atmospheres inside their thermodynamic seed formation window. This suggests that the atmospheres of exopla\-nets and brown dwarfs, as well as the optically thin parts of AGB outflows, may contain mainly monomers in the dynamic phase of cluster formation, but that a more uniform cluster abundance can be a fingerprint for the steady-state situation.

\cite{2000JPhB...33.3417J} provide the wavenumbers for the vibrational transitions for their DFT Gaussian94 (TiO$_2$)$_{\rm N}$ cluster simulations (DFT/B3P86/6-31G(d)) for cluster szies $N=1\,\ldots\,6,$ which may serve as a guide for observational approaches to disentangle the state of CCN formation in exoplanets, brown dwarfs, or also for AGB stars similar to \cite{2017A&A...608A..55D}. \cite{2000JPhB...33.3417J} point out that larger clusters of size $N$ have higher intensities for IR-active vibrations than smaller clusters. Although lower energy isomers of (TiO$_2$)$_N$, $N=3-6$, exist \citep{2014JCTC,2015A&A...575A..11L}, structural (geometric) arguments support our assumption that the cluster growth proceeds via the `gaiter'-shaped clusters derived by \cite{2000JPhB...33.3417J}. The timescale for a metastable isomer to relax to its global minimum structure can be long, in particular if the related potential energy surfaces (PES) consist of more than one funnel \citep{doye_1996}. The number of funnels for the relaxation of the (TiO$_2$)$_n$, $n=3-6$, clusters is unknown, since we did not perform a (computationally very demanding) scan of the PES. However, based on their atomic coordination, a relaxation of the (TiO$_2$)$_n$, $n=3-6$, 
clusters used in this study involves a re-arrangement (breaking and formation) of several bonds and is likely to proceed via several funnels.

The line opacity, $A_{\rm k}(N)\cdot f(N)$ [$10^{-8}$ cm$^{-1}$ s$^{-1}$],  for the integrated absorption coefficients data $A_{\rm k}$ [$10^{-8}$ cm$^2$s$^{-1}$], provided by \cite{JeongPhd} for $N=1\,\ldots\,6$, is shown in Fig.~\ref{fig:Ak_Jeong0}. The line opacity is evaluated at two different times: $t=10^7$s (top panel) when the cluster formation in very dynamic and for $t=10^9$s (bottom panel) when the cluster formation has reached a steady state (see Fig.~\ref{fig:MCTDcomp}). The line opacities are rather similar for both cases such that an observational differentiation between the two nucleation stages may be difficult. We note that cluster data are re-evaluated all the time; however, what ultimately determines the result is the number of clusters of a given size. The higher abundance of the smaller clusters outweighs the stronger IR line absorption cross sections of the larger clusters. It may however be feasible at some point to observe TiO$_2$ clusters and Fig.~\ref{fig:Ak_Jeong0} indicates that clusters $N=4, 5, 6$ would be the best candidates. We therefore reproduce the detailed vibrational absorption spectrum, $A_{\rm k}$ [$10^{-8}$ cm$^2$s$^{-1}$], for (TiO$_2$)$_{\rm 4}$, (TiO$_2$)$_{\rm 5}$  , and (TiO$_2$)$_{\rm 6}$ in Fig.~\ref{fig:Ak_Jeong} (data from Appendix C in \citealt{JeongPhd}). 

For each of these line spectra, the bottom right panel shows the Lorentz folded line spectra, which is constructed as follows. For given $N$, we construct the Lorentzian $L_{N,i}(\lambda)=C_i 2w/(4(\lambda-\lambda_{0,i})^2+w^2)$ for the wavelength $\lambda_{0,i}$ of the $i$-th peak, where we choose a fixed width of $w=1\ \mu$m and where $C_i$ is chosen such that the maximum of $L_{N,i}$ agrees with the maximum of the $i$-th line. Then for each $N$, the enveloped Lorentz profile is calculated as $L_{N}(\lambda)=\max\limits_{i} L_{N,i}(\lambda)$. As the total line spectrum depends on the particle number for each individual $N$, we calculate the total spectrum as the sum $L_4+L_5+L_6$ (dashed line).

The (TiO$_2$)$_{\rm 4}$, (TiO$_2$)$_{\rm 5}$  , and (TiO$_2$)$_{\rm 6}$  vibrational lines would be well covered by JWST/MIRI (Mid-Infrared Instrument) for $\lambda<45\mu$m or by ELT (Extremely Large Telescope) instruments like MIDIR (mid-IR imager). However, it may be difficult to disentangle individual cluster lines but instead be more feasible to detect a subset of lines of different clusters for example at the morning terminator of ultra-hot Jupiters. The morning terminator of ultra-hot Jupiters can be expected to be cloud free, hence optically thin to pressures of $p_{\rm gas}\approx 10^{-2}\ldots 10^{-4}$bar where the atmospheric temperature reaches the 1000K-window. In the future, the cross-correlation technique (e.g. \citealt{2019AJ....157..114B,2020AJ....160..198H,2020arXiv201110587S}) may be most suitable to explicitly search for cloud condensation clusters in exoplanet atmospheres if high-resolution opacity data for such clusters are available.



\section{Conclusions}\label{s:concl}

The study of (TiO$_2$)$_{\rm N}$ cluster formation presented here has used a 3D MC and a kinetic approach that is based on thermodynamic cluster data (kinetic nucleation). The 3D MC approach enables the following insights into  (TiO$_2$)$_{\rm N}$ cluster formation as key precursors for the formation of CCNs in exoplanet and brown dwarf  atmospheres, and in AGB star envelopes:
\begin{itemize}
    \item Modelling the formation of (TiO$_2$)$_{\rm N}$ clusters by monomer-cluster or by  cluster-cluster collisions  results in two fundamentally different cluster size distributions. The monomer-cluster scenario has a Gaussian-like shape with a well-defined mean cluster diameter. The cluster size distribution appears power-law like if cluster-cluster growth is enabled.
    \item Larger clusters form faster by cluster-cluster collisions instead of monomer-cluster collisions only.
    \item  A sweet-spot temperature occurs for most efficient cluster growth. This sweet-spot temperature value (1000K for TiO$_2$ \citep{2019MNRAS.489.4890B}) also depends on the gas density as both determine the collisional rates of the clusters. Therefore, cloud formation in exoplanet and brown dwarf atmospheres and the dust formation in AGB stars will be triggered within this temperature range.
    \item The cluster size distributions are different for different temperatures, which is driven by the interplay between growth and evaporation processes.
    \item We tested the effect of missing material data by studying different sticking probabilities. This material data insufficiency matters for the time evolution of the cluster size distributions but not for the final steady-state results.
    \item The onset of cloud (or dust) formation may be traced by observing  (TiO$_2$)$_{\rm 4}$, (TiO$_2$)$_{\rm 5}$  , and (TiO$_2$)$_{\rm 6}$  vibrational lines, which are present at temperatures around 1000 K and may be detectable with instruments such as JWST/MIRI ($\lambda<45\mu$m) or ELT/MIDIR. A dedicated search applying cross-correlation would be most desirable but requires a complete-as-possible,  high-resolution line list for each cluster.

\end{itemize}

\noindent
A comparison of results from the 3D MC method and the kinetic approach based on thermodynamic cluster data led to the following conclusions:
\begin{itemize}
\item The 3D MC approach enables the study of the time-resolved formation dynamics of the individual cluster growth, which is not accessible to the kinetic steady-state approach.
    \item For $T=1000$K, both methods agree well regarding cluster number densities for $N=1\,\ldots\,10$, the vivid onset of cluster formation, and the long transition into a steady state.
    \item The faster onset of cluster formation in the 3D MC compared to the kinetic results from the increased collisional rates in spots of increased gas density. A turbulent fluid field will cause similar effects.
    \item Our comparison also supports the 3D MC approach for H$_2$SO$_4$-H$_2$O  cluster formation by \cite{2018JCoPh.363...30K, 2020AerST..54.1007K} to model the early stages of CCN formation.
\end{itemize}


\begin{acknowledgements}
     Ch.H and M.B.E. acknowledge funding from the European Union H2020-MSCA-ITN-2019 under grant agreement no. 860470 (CHAMELEON), J.-P.~S. acknowledges funding from the St Andrews St Leonard College international scholarship. D.G. acknowledges support from the ERC consolidator grant 646758 AEROSOL. We thank H. Svensmark for discussions on the paper's topic.
\end{acknowledgements}


\bibliographystyle{aa}
\bibliography{reference.bib}

\begin{thebibliography}{59}
\expandafter\ifx\csname natexlab\endcsname\relax\def\natexlab#1{#1}\fi

\bibitem[{Barber {et~al.}(1996)Barber, Dobkin, \& Huhdanpaa}]{Barber1996}
Barber, C.~B., Dobkin, D.~P., \& Huhdanpaa, H. 1996, ACM Transactions on
  Mathematical Software, 22, 469

\bibitem[{{Barstow}(2020)}]{2020MNRAS.497.4183B}
{Barstow}, J.~K. 2020, \mnras, 497, 4183

\bibitem[{{Berardo} {et~al.}(2014){Berardo}, {Hu}, {Shevlin}, {Woodley},
  {Kowalski}, \& {Zwijnenburg}}]{2014JCTC}
{Berardo}, E., {Hu}, H.-S., {Shevlin}, S.~A., {et~al.} 2014, Journal of
  Chemical Theory and Computation, 10, 1189

\bibitem[{{Boulangier} {et~al.}(2019){Boulangier}, {Gobrecht}, {Decin}, {de
  Koter}, \& {Yates}}]{2019MNRAS.489.4890B}
{Boulangier}, J., {Gobrecht}, D., {Decin}, L., {de Koter}, A., \& {Yates}, J.
  2019, \mnras, 489, 4890

\bibitem[{{Brogi} \& {Line}(2019)}]{2019AJ....157..114B}
{Brogi}, M. \& {Line}, M.~R. 2019, \aj, 157, 114

\bibitem[{{Br{\"u}nken} {et~al.}(2008){Br{\"u}nken}, {M{\"u}ller}, {Menten},
  {McCarthy}, \& {Thaddeus}}]{bruenken_2008}
{Br{\"u}nken}, S., {M{\"u}ller}, H., {Menten}, K., {McCarthy}, M., \&
  {Thaddeus}, P. 2008, \aap, 676, 1367

\bibitem[{{Chang} {et~al.}(2013){Chang}, {Patzer}, {Kegel}, \&
  {Chandra}}]{2013Ap&SS.347..315C}
{Chang}, C., {Patzer}, A.~B.~C., {Kegel}, W.~H., \& {Chandra}, S. 2013, \apss,
  347, 315

\bibitem[{{Chang} {et~al.}(2005){Chang}, {Patzer}, {Sedlmayr}, \&
  {S{\"u}lzle}}]{2005PhRvB..72w5402C}
{Chang}, C., {Patzer}, A.~B.~C., {Sedlmayr}, E., \& {S{\"u}lzle}, D. 2005,
  \prb, 72, 235402

\bibitem[{{Col{\'o}n} {et~al.}(2020){Col{\'o}n}, {Kreidberg}, {Welbanks},
  {Line}, {Madhusudhan}, {Beatty}, {Tamburo}, {Stevenson}, {Mandell},
  {Rodriguez}, {Barclay}, {Lopez}, {Stassun}, {Angerhausen}, {Fortney},
  {James}, {Pepper}, {Ahlers}, {Plavchan}, {Awiphan}, {Kotnik}, {McLeod},
  {Murawski}, {Chotani}, {LeBrun}, {Matzko}, {Rea}, {Vidaurri}, {Webster},
  {Williams}, {Cox}, {Tan}, \& {Gilbert}}]{2020AJ....160..280C}
{Col{\'o}n}, K.~D., {Kreidberg}, L., {Welbanks}, L., {et~al.} 2020, \aj, 160,
  280

\bibitem[{{Decin} {et~al.}(2017){Decin}, {Richards}, {Waters}, {Danilovich},
  {Gobrecht}, {Khouri}, {Homan}, {Bakker}, {Van de Sande}, {Nuth}, \& {De
  Beck}}]{2017A&A...608A..55D}
{Decin}, L., {Richards}, A.~M.~S., {Waters}, L.~B.~F.~M., {et~al.} 2017, \aap,
  608, A55

\bibitem[{{Doye} \& {Wales}(1996)}]{doye_1996}
{Doye}, J. \& {Wales}, J. 1996, J. Chem. Phys., 105, 8428

\bibitem[{Dunne {et~al.}(2016)Dunne, Gordon, K{\"u}rten, Almeida, Duplissy,
  Williamson, Ortega, Pringle, Adamov, Baltensperger, Barmet, Benduhn, Bianchi,
  Breitenlechner, Clarke, Curtius, Dommen, Donahue, Ehrhart, Flagan, Franchin,
  Guida, Hakala, Hansel, Heinritzi, Jokinen, Kangasluoma, Kirkby, Kulmala,
  Kupc, Lawler, Lehtipalo, Makhmutov, Mann, Mathot, Merikanto, Miettinen,
  Nenes, Onnela, Rap, Reddington, Riccobono, Richards, Rissanen, Rondo,
  Sarnela, Schobesberger, Sengupta, Simon, Sipil{\"a}, Smith, Stozkhov,
  Tom{\'e}, Tr{\"o}stl, Wagner, Wimmer, Winkler, Worsnop, \&
  Carslaw}]{Dunne2016}
Dunne, E.~M., Gordon, H., K{\"u}rten, A., {et~al.} 2016, Science, 354, 1119

\bibitem[{{Gobrecht} {et~al.}(2018){Gobrecht}, {Decin}, {Cristallo}, \&
  {Bromley}}]{gobrecht_2018}
{Gobrecht}, D., {Decin}, L., {Cristallo}, S., \& {Bromley}, S. 2018, Chem.
  Phys. Lett., 711, 138

\bibitem[{{Helling}(2020)}]{2020arXiv201103302H}
{Helling}, C. 2020, arXiv e-prints, arXiv:2011.03302

\bibitem[{{Helling} \& {Fomins}(2013)}]{2013RSPTA.37110581H}
{Helling}, C. \& {Fomins}, A. 2013, Philosophical Transactions of the Royal
  Society of London Series A, 371, 20110581

\bibitem[{{Helling} {et~al.}(2019){Helling}, {Iro}, {Corrales}, {Samra},
  {Ohno}, {Alam}, {Steinrueck}, {Lew}, {Molaverdikhani}, {MacDonald},
  {Herbort}, {Woitke}, \& {Parmentier}}]{2019A&A...631A..79H}
{Helling}, C., {Iro}, N., {Corrales}, L., {et~al.} 2019, \aap, 631, A79

\bibitem[{{Helling} {et~al.}(2004){Helling}, {Klein}, {Woitke}, {Nowak}, \&
  {Sedlmayr}}]{2004A&A...423..657H}
{Helling}, C., {Klein}, R., {Woitke}, P., {Nowak}, U., \& {Sedlmayr}, E. 2004,
  \aap, 423, 657

\bibitem[{{Helling} {et~al.}(2017){Helling}, {Tootill}, {Woitke}, \&
  {Lee}}]{2017A&A...603A.123H}
{Helling}, C., {Tootill}, D., {Woitke}, P., \& {Lee}, G. 2017, \aap, 603, A123

\bibitem[{{Helling} \& {Woitke}(2006)}]{2006A&A...455..325H}
{Helling}, C. \& {Woitke}, P. 2006, \aap, 455, 325

\bibitem[{{Hood} {et~al.}(2020){Hood}, {Fortney}, {Line}, {Martin}, {Morley},
  {Birkby}, {Rustamkulov}, {Lupu}, \& {Freedman}}]{2020AJ....160..198H}
{Hood}, C.~E., {Fortney}, J.~J., {Line}, M.~R., {et~al.} 2020, \aj, 160, 198

\bibitem[{Jeong(2000)}]{JeongPhd}
Jeong, K.~S. 2000, Dust shells around oxygen-rich Miras and long-period
  variables (TU Berlin, PhD Thesis)

\bibitem[{{Jeong} {et~al.}(2000){Jeong}, {Chang}, {Sedlmayr}, \&
  {S{\"u}lzle}}]{2000JPhB...33.3417J}
{Jeong}, K.~S., {Chang}, C., {Sedlmayr}, E., \& {S{\"u}lzle}, D. 2000, Journal
  of Physics B Atomic Molecular Physics, 33, 3417

\bibitem[{Koch \& Manzhos(2017)}]{Koch2017}
Koch, D. \& Manzhos, S. 2017, Journal of Physical Chemistry Letters, 8, 1593

\bibitem[{{K{\"o}hn} {et~al.}(2018){K{\"o}hn}, {Enghoff}, \&
  {Svensmark}}]{2018JCoPh.363...30K}
{K{\"o}hn}, C., {Enghoff}, M.~B., \& {Svensmark}, H. 2018, Journal of
  Computational Physics, 363, 30

\bibitem[{{K{\"o}hn} {et~al.}(2020){K{\"o}hn}, {Enghoff}, \&
  {Svensmark}}]{2020AerST..54.1007K}
{K{\"o}hn}, C., {Enghoff}, M.~B., \& {Svensmark}, H. 2020, Aerosol Science
  Technology, 54, 1007

\bibitem[{{Lacy} \& {Burrows}(2020)}]{2020ApJ...904...25L}
{Lacy}, B.~I. \& {Burrows}, A. 2020, \apj, 904, 25

\bibitem[{{Lam} {et~al.}(2015){Lam}, {Amans}, {Dujardin}, {Ledoux}, \&
  {Allouche}}]{2015JPCA..119.8944L}
{Lam}, J., {Amans}, D., {Dujardin}, C., {Ledoux}, G., \& {Allouche}, A.-R.
  2015, Journal of Physical Chemistry A, 119, 8944

\bibitem[{{Lee} {et~al.}(2016){Lee}, {Dobbs-Dixon}, {Helling}, {Bognar}, \&
  {Woitke}}]{2016A&A...594A..48L}
{Lee}, G., {Dobbs-Dixon}, I., {Helling}, C., {Bognar}, K., \& {Woitke}, P.
  2016, \aap, 594, A48

\bibitem[{{Lee} {et~al.}(2015){Lee}, {Helling}, {Giles}, \&
  {Bromley}}]{2015A&A...575A..11L}
{Lee}, G., {Helling}, C., {Giles}, H., \& {Bromley}, S.~T. 2015, \aap, 575, A11

\bibitem[{{Lee} {et~al.}(2018){Lee}, {Blecic}, \&
  {Helling}}]{2018A&A...614A.126L}
{Lee}, G.~K.~H., {Blecic}, J., \& {Helling}, C. 2018, \aap, 614, A126

\bibitem[{{Lines} {et~al.}(2018){Lines}, {Mayne}, {Boutle}, {Manners}, {Lee},
  {Helling}, {Drummond}, {Amundsen}, {Goyal}, {Acreman}, {Tremblin}, \&
  {Kerslake}}]{2018A&A...615A..97L}
{Lines}, S., {Mayne}, N.~J., {Boutle}, I.~A., {et~al.} 2018, \aap, 615, A97

\bibitem[{{Lines} {et~al.}(2019){Lines}, {Mayne}, {Manners}, {Boutle},
  {Drummond}, {Mikal-Evans}, {Kohary}, \& {Sing}}]{2019MNRAS.488.1332L}
{Lines}, S., {Mayne}, N.~J., {Manners}, J., {et~al.} 2019, \mnras, 488, 1332

\bibitem[{{Nikolov} {et~al.}(2018){Nikolov}, {Sing}, {Fortney}, {Goyal},
  {Drummond}, {Evans}, {Gibson}, {De Mooij}, {Rustamkulov}, {Wakeford},
  {Smalley}, {Burgasser}, {Hellier}, {Helling}, {Mayne}, {Madhusudhan},
  {Kataria}, {Baines}, {Carter}, {Ballester}, {Barstow}, {McCleery}, \&
  {Spake}}]{2018Natur.557..526N}
{Nikolov}, N., {Sing}, D.~K., {Fortney}, J.~J., {et~al.} 2018, \nat, 557, 526

\bibitem[{Olenius {et~al.}(2018)Olenius, Pichelstorfer, Stolzenburg, Winkler,
  Lehtinen, \& Riipinen}]{olenius2018}
Olenius, T., Pichelstorfer, L., Stolzenburg, D., {et~al.} 2018, Sci. Rep., 8,
  14160

\bibitem[{{Ormel} \& {Min}(2019)}]{2019A&A...622A.121O}
{Ormel}, C.~W. \& {Min}, M. 2019, \aap, 622, A121

\bibitem[{{Parmentier} {et~al.}(2020){Parmentier}, {Showman}, \&
  {Fortney}}]{2020MNRAS.tmp.3310P}
{Parmentier}, V., {Showman}, A.~P., \& {Fortney}, J.~J. 2020, \mnras, 501, 78

\bibitem[{{Patzer} {et~al.}(1999){Patzer}, {Chang}, {Sedlmayr}, \&
  {S{\"u}lzle}}]{1999EPJD....6...57P}
{Patzer}, A.~B.~C., {Chang}, C., {Sedlmayr}, E., \& {S{\"u}lzle}, D. 1999,
  European Physical Journal D, 6, 57

\bibitem[{{Patzer} {et~al.}(2005){Patzer}, {Chang}, {Sedlmayr}, \&
  {S{\"u}lzle}}]{2005EPJD...32..329P}
{Patzer}, A.~B.~C., {Chang}, C., {Sedlmayr}, E., \& {S{\"u}lzle}, D. 2005,
  European Physical Journal D, 32, 329

\bibitem[{{Patzer} {et~al.}(2014){Patzer}, {Chang}, \&
  {S{\"u}lzle}}]{2014CPL...612...39P}
{Patzer}, A.~B.~C., {Chang}, C., \& {S{\"u}lzle}, D. 2014, Chemical Physics
  Letters, 612, 39

\bibitem[{{Patzer} {et~al.}(1998){Patzer}, {Gauger}, \&
  {Sedlmayr}}]{1998A&A...337..847P}
{Patzer}, A.~B.~C., {Gauger}, A., \& {Sedlmayr}, E. 1998, \aap, 337, 847

\bibitem[{{Powell} {et~al.}(2018){Powell}, {Zhang}, {Gao}, \&
  {Parmentier}}]{2018ApJ...860...18P}
{Powell}, D., {Zhang}, X., {Gao}, P., \& {Parmentier}, V. 2018, \apj, 860, 18

\bibitem[{{Roman} {et~al.}(2021){Roman}, {Kempton}, {Rauscher}, {Harada},
  {Bean}, \& {Stevenson}}]{2020arXiv201006936R}
{Roman}, M.~T., {Kempton}, E. M.~R., {Rauscher}, E., {et~al.} 2021, Astro. J.,
  908, 101

\bibitem[{{Samra} {et~al.}(2020){Samra}, {Helling}, \&
  {Min}}]{2020A&A...639A.107S}
{Samra}, D., {Helling}, C., \& {Min}, M. 2020, \aap, 639, A107

\bibitem[{{Schmidt} {et~al.}(2005){Schmidt}, {Hillebrandt}, \&
  {Niemeyer}}]{2005CTM.....9..693S}
{Schmidt}, W., {Hillebrandt}, W., \& {Niemeyer}, J.~C. 2005, Combustion Theory
  and Modelling, 9, 693

\bibitem[{{Sedlmayr}(1994)}]{1994LNP...428..163S}
{Sedlmayr}, E. 1994, in Lecture Notes in Physics, Berlin Springer Verlag, Vol.
  428, IAU Colloq. 146: Molecules in the Stellar Environment, ed. U.~G.
  {Jorgensen}, 163

\bibitem[{Seinfeld \& Pandis(2006)}]{seinfeld_2006}
Seinfeld, J. \& Pandis, S. 2006, Atmospheric Chemistry and Physics (John Wiley
  \& Sons, New Jersey)

\bibitem[{{Serindag} {et~al.}(2020){Serindag}, {Nugroho}, {Molli{\`e}re}, {de
  Mooij}, {Gibson}, \& {Snellen}}]{2020arXiv201110587S}
{Serindag}, D.~B., {Nugroho}, S.~K., {Molli{\`e}re}, P., {et~al.} 2020, Astro.
  \& Astro., 645, A90

\bibitem[{{Spake} {et~al.}(2020){Spake}, {Sing}, {Wakeford}, {Nikolov},
  {Mikal-Evans}, {Deming}, {Barstow}, {Anderson}, {Carter}, {Gillon}, {Goyal},
  {Hebrard}, {Hellier}, {Kataria}, {Lam}, {Triaud}, \&
  {Wheatley}}]{2020MNRAS.tmp.2941S}
{Spake}, J.~J., {Sing}, D.~K., {Wakeford}, H.~R., {et~al.} 2020, \mnras, 500,
  4042

\bibitem[{Stocker {et~al.}(2013)Stocker, Qin, Plattner, Tignor, Allen,
  Boschung, Nauels, Xia, Bex, \& Midgley}]{IPCCAR5SPM}
Stocker, T., Qin, D., Plattner, G.-K., {et~al.} 2013, IPCC, 2013: Summary for
  Policymakers. In: Climate Change 2013: The Physical Science
  Basis.Contribution of Working Group I to the Fifth Assessment Report of the
  Intergovernmental Panel on Climate Change (Cambridge University Press)

\bibitem[{{Svensmark} {et~al.}(2013){Svensmark}, {Enghoff}, \&
  {Pedersen}}]{2013PhLA..377.2343S}
{Svensmark}, H., {Enghoff}, M.~B., \& {Pedersen}, J. O.~P. 2013, Physics
  Letters A, 377, 2343

\bibitem[{{Svensmark} {et~al.}(2017){Svensmark}, {Enghoff}, {Shaviv}, \&
  {Svensmark}}]{2017NatCo...8.2199S}
{Svensmark}, H., {Enghoff}, M.~B., {Shaviv}, N.~J., \& {Svensmark}, J. 2017,
  Nature Communications, 8, 2199

\bibitem[{{Svensmark} {et~al.}(2016){Svensmark}, {Enghoff}, {Shaviv}, \&
  {Svensmark}}]{2016JGRA..121.8152S}
{Svensmark}, J., {Enghoff}, M.~B., {Shaviv}, N.~J., \& {Svensmark}, H. 2016,
  Journal of Geophysical Research (Space Physics), 121, 8152

\bibitem[{{Svensmark} {et~al.}(2020){Svensmark}, {Shaviv}, {Enghoff}, \&
  {Svensmark}}]{2020E&SS....701142S}
{Svensmark}, J., {Shaviv}, N.~J., {Enghoff}, M.~B., \& {Svensmark}, H. 2020,
  Earth and Space Science, 7, e01142

\bibitem[{{Witte} {et~al.}(2011){Witte}, {Helling}, {Barman}, {Heidrich}, \&
  {Hauschildt}}]{2011A&A...529A..44W}
{Witte}, S., {Helling}, C., {Barman}, T., {Heidrich}, N., \& {Hauschildt},
  P.~H. 2011, \aap, 529, A44

\bibitem[{{Witte} {et~al.}(2009){Witte}, {Helling}, \&
  {Hauschildt}}]{2009A&A...506.1367W}
{Witte}, S., {Helling}, C., \& {Hauschildt}, P.~H. 2009, \aap, 506, 1367

\bibitem[{{Woitke} \& {Helling}(2003)}]{2003A&A...399..297W}
{Woitke}, P. \& {Helling}, C. 2003, \aap, 399, 297

\bibitem[{{Woitke} \& {Helling}(2004)}]{2004A&A...414..335W}
{Woitke}, P. \& {Helling}, C. 2004, \aap, 414, 335

\bibitem[{{Woitke} {et~al.}(2018){Woitke}, {Helling}, {Hunter}, {Millard},
  {Turner}, {Worters}, {Blecic}, \& J.W.}]{woitke_2018}
{Woitke}, P., {Helling}, C., {Hunter}, G., {et~al.} 2018, \aap, 614, A1

\bibitem[{Yu(2005)}]{yu_2005}
Yu, F. 2005, J. Chem. Phys., 122, 074501

\end{thebibliography}

\newpage
\appendix

\section{Supplementary figures}

\begin{figure*}
    \centering
    \includegraphics[scale=0.55]{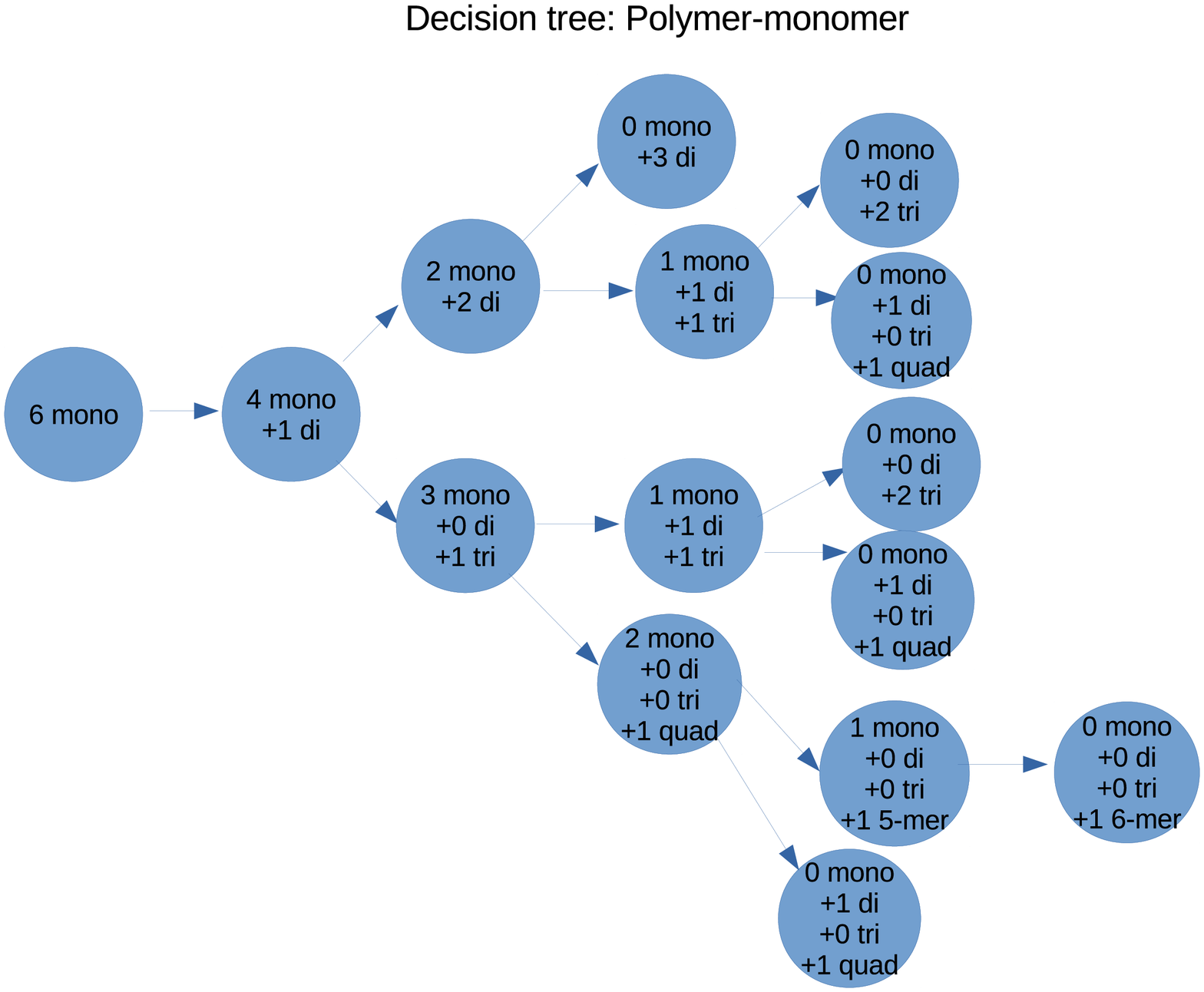}
 \includegraphics[scale=0.55]{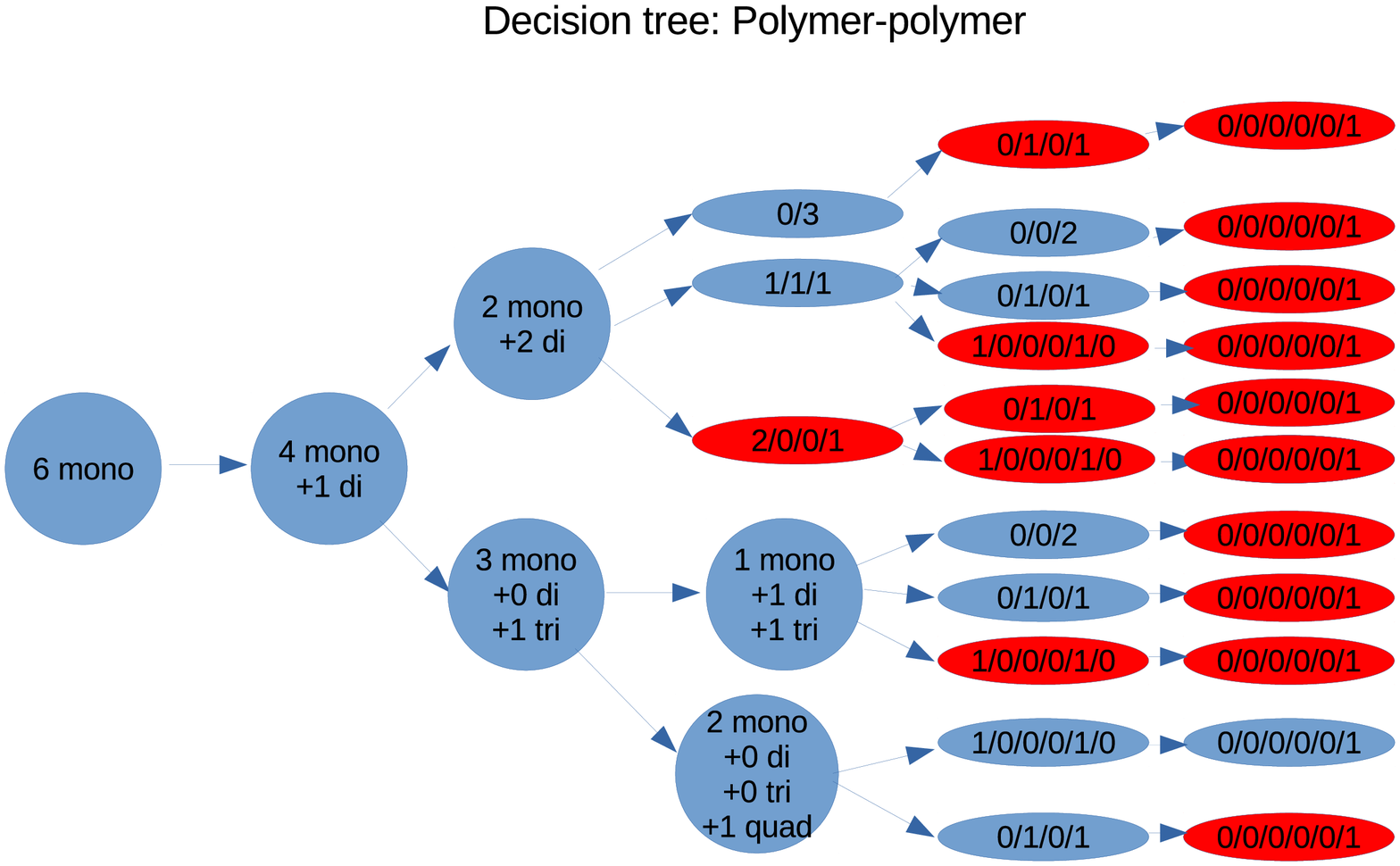}
    \caption{Decision tree of how six initial monomers can grow. {\bf Top}: Without cluster-cluster growth. {\bf Bottom}: With cluster-cluster growth allowed. In red we show those paths that do not appear in the polymer-monomer case. The slash notation in the bottom panel gives the individual number of $N$-mers ordered by their position in the slash line. For polymer-polymer growth, all possible paths end up with one single 6-mer.}
    \label{fig:dectre}
\end{figure*}

\end{document}